\documentclass[twocolumn,showpacs,preprintnumbers,amsmath,amssymb,epsfig]{revtex4}

\usepackage{graphicx}
\usepackage{dcolumn}
\usepackage{bm}
\usepackage{epsfig}

\usepackage{float}
\usepackage{amsmath}
\usepackage{epsfig,floatflt}
\usepackage{subfigure}
\usepackage[usenames]{color}

\begin{document}


\title{Using SPTpol, Planck 2015, and non-CMB data to constrain tilted spatially-flat and untilted non-flat $\Lambda$CDM, XCDM, and $\phi$CDM dark energy inflation cosmologies}

\author{Chan-Gyung Park${}^{1}$ and Bharat Ratra${}^{2}$}
\affiliation{
         ${}^{1}$Division of Science Education and Institute of Fusion
                 Science, Jeonbuk National University, Jeonju 54896, Republic of Korea  \\
         ${}^{2}$Department of Physics, Kansas State University, 116 Cardwell Hall,
                 Manhattan, KS 66506, USA}
\email{park.chan.gyung@gmail.com,ratra@phys.ksu.edu}
\date{\today}


\begin{abstract}
We use six tilted spatially-flat and untilted non-flat dark energy cosmological models in analyses of South Pole Telescope polarization (SPTpol) cosmic microwave background (CMB) data, alone and in combination with Planck 2015 CMB data and non-CMB data, namely, the Pantheon Type Ia supernovae apparent magnitudes, a collection of baryon acoustic oscillation data points, Hubble parameter measurements, and growth rates.
Although the cosmological models that best-fit the Planck CMB and non-CMB data do not provide good fits to the SPTpol data, with the $\chi^2$'s exceeding the expected value, given the uncertainties, in each model the cosmological parameter constraints from the SPTpol data and from the Planck CMB and non-CMB data are largely mutually consistent.
When the smaller angular scale SPTpol data are used jointly with either the Planck data alone or with the Planck CMB and the non-CMB data to constrain untilted non-flat models, spatially-closed models remain favored over their corresponding flat limits. When used in conjunction with Planck data, non-CMB data (baryon acoustic oscillation measurements in particular, from six experiments) have significantly more constraining power than the SPTpol data.
\end{abstract}

\maketitle

\section{Introduction}
\label{sec:intro}

A main goal of cosmology research is to use astronomical observations to 
measure the parameters of the cosmological model as accurately as possible.
The currently widely accepted standard model is the spatially-flat 
$\Lambda$CDM model, \cite{Peebles1984}, where the cosmological constant 
($\Lambda$) dark energy and the cold dark matter (CDM) constitute 95\% of 
the present energy content of the Universe with baryonic matter 
contributing the remaining 5\% and the structure of the Universe has grown 
under gravitational instability from primordial infinitesimal 
quantum-mechanical energy density perturbations generated 
during the very early epoch of near slow-roll inflation. The main observations 
that lend significant support to this standard model include CMB anisotropy 
data \cite{PlanckCollaboration2016,PlanckCollaboration2018}, 
Type Ia supernovae redshift-magnitude data \cite{Scolnicetal2018}, baryonic 
acoustic oscillations (BAO) data \cite{Alametal2017,Beutleretal2011,Rossetal2015,Ataetal2018,Bautistaetal2017,Font-Riberaetal2014}, and Hubble parameter 
$[H(z)]$ measurements \cite{Morescoetal2016,Farooqetal2017}.

Although the spatially-flat model with a time-independent cosmological constant as dark energy 
is widely accepted, the possibility still remains that space is not flat and 
that the dark energy is dynamical. Recent joint analyses of the Planck 2015 CMB 
and non-CMB observations in the untilted non-flat $\Lambda$CDM model show
evidence for spatial non-flatness with $5.2\sigma$ significance 
\cite{Oobaetal2018a,ParkRatra2018a,ParkRatra2018b}. Here the untilted non-flat 
model does not have the density perturbation power spectral index $n_s$ as a 
free parameter, unlike in the commonly used tilted spatially-flat model. 
Also, the Planck CMB 
and non-CMB observational data do not rule out the possibility of dark energy 
being dynamical, \cite{Oobaetal2018d,ParkRatra2018b,ParkRatra2018c,DDEsupport}.
Recent research also demonstrates that non-flat dynamical dark energy models 
with constant dark energy equation of state (XCDM) or based on a  minimally 
coupled scalar field ($\phi$CDM) \cite{PR1988} are still allowed by the 
current observations
\cite{Oobaetal2018b,Oobaetal2018c,ParkRatra2018b,ParkRatra2018c,SpatialCurvature}. In the XCDM parameterization the dark energy density decreases with time. Here the dark energy is modeled as an ideal $X$-fluid with equation of state $p_X = w \rho_X$ where $p_X$ and $\rho_X$ are the fluid pressure and energy density and the equation of state parameter $w < - 1/3$. Since this is a widely-used parameterization, we also consider it here. However, it is incomplete and needs to be extended since it is unable to consistently describe the evolution of spatial inhomogeneities. On the other hand, the $\phi$CDM model --- in which a scalar field is used to model dynamical dark energy --- is physically complete and consistent.

While Planck CMB data have been widely used to constrain cosmological models,
there also exist several CMB data sets from higher-resolution ground-based 
CMB observatories such as the Atacama Cosmology Telescope (ACT; 
\cite{Dasetal2014}) and the South Pole Telescope (SPT; \cite{SPT}). For 
example, the South Pole Telescope provides information on the CMB temperature 
and polarization spatial anisotropies at angular scales much smaller than the 
range probed by the Planck satellite (SPTpol; \cite{Henningetal2018}). It 
is essential to investigate whether the Planck CMB data and other high 
resolution observations are mutually consistent. There have been two 
comparisons of the cosmological constraints from the Planck data and the SPTpol 
data, both of which made use of the tilted flat $\Lambda$CDM model. The 
SPTpol collaboration, \cite{Henningetal2018}, concluded that the 
Planck 2015 and SPTpol results are mildly inconsistent in this model, while 
the Planck collaboration, \cite{PlanckCollaboration2018}, concluded that 
the Planck 2018 and SPTpol results are not inconsistent. However, no such 
consistency check has been made for other simple cosmological models, such as 
the flat dynamical dark energy (XCDM and $\phi$CDM) models and the non-flat 
$\Lambda$CDM, XCDM, and $\phi$CDM models, so it is not known if the above results 
are model independent. 

Here we constrain cosmological parameters in the tilted flat and untilted 
non-flat $\Lambda$CDM, XCDM, and $\phi$CDM dark energy models using the 
recent SPTpol CMB data sets, and study the consistency between the best-fit 
models favored by the SPTpol data and those favored by the Planck 2015 CMB 
and the non-CMB data \cite{ParkRatra2018b,ParkRatra2018c}.
The non-CMB data used in our analysis includes the Pantheon Type Ia supernovae apparent magnitudes \cite{Scolnicetal2018}, a collection of BAO data points \cite{Alametal2017,Beutleretal2011,Rossetal2015,Ataetal2018,Bautistaetal2017,Font-Riberaetal2014}, Hubble parameter measurements, and growth rates \cite{ParkRatra2018a}.
Using a variety of cosmological models to compare different data sets has proved instructive, 
allowing somewhat model-independent conclusions to be drawn, 
\cite{Ratraetal1999}. We find that the constraints on model parameters from the SPTpol data and from the Planck CMB and the non-CMB data are not inconsistent and so we also derive constraints on cosmological parameters from joint analyses of all data.  

The main ``model-independent'' conclusions of our analyses are: (i) All Planck CMB and non-CMB data best-fit cosmological models we study do not provide good fits to the SPTpol data.
(ii) However, there is mutual consistency between the parameter constraints for a model that best fits the full SPTpol data and those for the same model that best fits the Planck CMB and the non-CMB data. 
(iii) When the full SPTpol data are used in joint 
analyses with either the Planck data alone or with the Planck CMB and 
the non-CMB data to constrain untilted non-flat models, closed models 
continue to be favored over their spatially-flat limits. (iv) When used  
together with Planck data, non-CMB data (BAO data in particular, from six 
experiments) have 
significantly more constraining power than do the SPTpol data. 

This paper is organized as follows. Section II describes the observational 
data used in our analyses while Sec.\ III describes our method for 
constraining cosmological parameters of the six tilted flat and untilted 
non-flat $\Lambda$CDM, XCDM, and $\phi$CDM models using various combinations
of data sets. The inflation model power spectra, that define the cosmological
models we use, are discussed in Sec.\ IV. Results are presented in Sec.\ V 
and a summary is provided in Sec. VI.

\section{Data}
\label{sec:data}

In this work, recent CMB and non-CMB data are used to constrain the tilted 
flat and untilted non-flat $\Lambda$CDM, XCDM, and $\phi$CDM dark energy models.
We use exactly the same Planck CMB and non-CMB data sets used in Refs.\ \cite{ParkRatra2018b,ParkRatra2018c}
for consistency with the previous works and to see what happens when the SPTpol data set is added to the mix.

We use the SPTpol and Planck 2015 CMB anisotropy data sets. The SPTpol data 
is composed
of the CMB $E$-mode polarization angular power spectrum (EE) and the temperature-$E$-mode cross-power spectrum (TE) over the spherical
harmonic multipoles $50 < \ell \le 8000$, based on CMB temperature and polarization observations on 500 $\textrm{deg}^2$ of the sky
\cite{Henningetal2018}. Here we use three different SPTpol data sets, namely, TE+EE, TE, and EE band power measurements, and the corresponding
covariance matrices, which are all publicly available at the South Pole Telescope website.

We also use the Planck 2015 CMB anisotropy data (TT+lowP) including the lensing data \cite{PlanckCollaboration2016},
where TT denotes the CMB temperature angular power spectrum at low $\ell$ ($2 \le \ell \le 29$) and
high $\ell$ ($30 \le \ell \le 2508$) and lowP represents the low-$\ell$ CMB TE, EE, and BB polarization
power spectra at $2 \le \ell \le 29$.
The primary reason for using Planck TT+lowP CMB data is that TT+lowP data was the baseline power spectrum data for Planck 2015 analysis and that the same data was used in the SPTpol group analysis. One may think that the Planck CMB data with polarization at high $\ell$ should be used for fair comparison with the STPpol data that span a wide range of $\ell$. The parameters of the flat $\Lambda$CDM model constrained by Planck TT+lowP and TT,TE,EE+lowP data sets (with high-$\ell$ polarization information) are totally consistent with each other with deviation less than $0.2\sigma$ (see Table 3 of Ref.\ \cite{PlanckCollaboration2016}). For the nonflat $\Lambda$CDM model, the devitation is less than $0.6\sigma$ \cite{PlanckParameterTables2015}. Thus, it is valid to use the TT+lowP baseline power spectrum data when comparing with SPTpol data. 
 
To get tighter constraints on model parameters, we jointly use the Planck and 
SPTpol data, together with non-CMB data. The non-CMB data sets we use are the 
Pantheon supernova Type Ia measurements (SN) \cite{Scolnicetal2018}, a compilation of BAO data \cite{Alametal2017,Beutleretal2011,Rossetal2015,Ataetal2018,Bautistaetal2017,Font-Riberaetal2014}, $H(z)$ measurements, and growth rates ($f\sigma_8$) (see Refs.\ \cite{ParkRatra2018b,ParkRatra2018c} for detailed description of these data).

\begin{table}
\caption{BAO measurements.}
\begin{ruledtabular}
\begin{tabular}{ccc}
 $z_\textrm{eff}$                     &  Measurement                                          &   Reference    \\[+0mm]
 \hline \\[-2mm]
 $0.38$     & $D_M (r_{d,\textrm{fid}} / r_d)$ [Mpc]                   $= 1512.39 \pm24.99$   &  \cite{Alametal2017} \\[+1mm]
 $0.38$     & $H (r_d / r_{d,\textrm{fid}})$ [km s$^{-1}$ Mpc$^{-1}$]  $= 81.21   \pm 2.37$   &  \cite{Alametal2017}  \\[+1mm]
 $0.51$     & $D_M (r_{d,\textrm{fid}} / r_d)$ [Mpc]                   $= 1975.22 \pm 30.10$  &  \cite{Alametal2017}  \\[+1mm]
 $0.51$     & $H (r_d / r_{d,\textrm{fid}})$ [km s$^{-1}$ Mpc$^{-1}$]  $= 90.90   \pm 2.33$   &  \cite{Alametal2017}  \\[+1mm]
 $0.61$     & $D_M (r_{d,\textrm{fid}} / r_d)$ [Mpc]                   $= 2306.68 \pm 37.08$  &  \cite{Alametal2017}  \\[+1mm]
 $0.61$     & $H (r_d / r_{d,\textrm{fid}})$ [km s$^{-1}$ Mpc$^{-1}$]  $= 98.96   \pm 2.50$   &  \cite{Alametal2017}  \\[+1mm]
 $0.38$     & $f \sigma_8                                               =0.497\pm0.045$       &  \cite{Alametal2017}  \\[+1mm]
 $0.51$     & $f \sigma_8                                               =0.458\pm0.038$       &  \cite{Alametal2017}  \\[+1mm]
 $0.61$     & $f \sigma_8                                               =0.436\pm0.034$       &  \cite{Alametal2017}  \\[+1mm]
 \hline \\[-2mm]
 $0.106$    & $r_d / D_V$                                              $= 0.327\pm0.015$  &  \cite{Beutleretal2011} \\[+1mm]
  \hline \\[-2mm]
 $0.15$     & $D_V (r_{d,\textrm{fid}} / r_d)$ [Mpc]                   $= 664\pm25$       & \cite{Rossetal2015}  \\[+1mm]
  \hline \\[-2mm]
 $1.52$     & $D_V (r_{d,\textrm{fid}} / r_d)$ [Mpc]                   $= 3843\pm147$     & \cite{Ataetal2018}  \\[+1mm]
   \hline \\[-2mm]
 $2.33$    &   $D_H^{0.7} D_M^{0.3} / r_d$                             $= 13.94\pm0.35$   & \cite{Bautistaetal2017} \\[+1mm]
   \hline \\[-2mm]
 $2.36$    & $D_H / r_d$                                               $= 9.0\pm0.3$      & \cite{Font-Riberaetal2014} \\[+1mm]
 $2.36$    & $D_A / r_d$                                               $= 10.8\pm0.4$     & \cite{Font-Riberaetal2014} \\[+0mm]
\end{tabular}
\\[+1mm]
Note: The sound horizon size (at the drag epoch) of the fiducial model
is $r_{d,\textrm{fid}}=147.78~\textrm{Mpc}$ in \cite{Alametal2017} and \cite{Ataetal2018},
and $r_{d,\textrm{fid}}=148.69~\textrm{Mpc}$ in \cite{Rossetal2015}.
\end{ruledtabular}
\label{tab:bao}
\end{table}

In Table \ref{tab:bao}, we summarize the BAO data points used in our analysis. Our collection encloses the BAO data sets adopted by the recent Planck 2018 data analysis \cite{PlanckCollaboration2018}. Here $D_M(z)$ and $H(z)$ are the comoving distance and Hubble parameter at redshift $z$, respectively, $D_V(z) =[cz D^2_M(z)/H(z)]^{1/3}$, $D_H(z)=c/H(z)$ with $c$ the speed of light, $r_d$ the sound horizon size at the drag epoch ($z_d$), and $f\sigma_8$ the growth rate. Note that the BAO data set of Ref.\ \cite{Alametal2017} includes Hubble parameters and growth rates as well as distance information and that  in the actual analysis for the data points of Ref.\ \cite{Rossetal2015,Font-Riberaetal2014} the probability distributions instead of the approximate Gaussian constraints are used (see Sec.\ 2.3 of Ref.\ \cite{ParkRatra2018a} for a more detailed description).

\section{Methods}

We apply the Markov chain Monte Carlo (MCMC) method, implemented in a 
modified version of the CAMB/COSMOMC program (version of Nov.\ 2016) 
\cite{CAMBCOSMOMC}, to explore the parameter space of the dark energy 
inflation models. The CAMB program computes the matter and CMB power 
spectra based
on the evolution of density perturbations of matter and radiation components and the COSMOMC program
estimates the parameter constraints that are favored by the given observational data sets using the MCMC method.

The tilted flat $\Lambda$CDM model is characterized by six cosmological parameters ($\Omega_b h^2$, $\Omega_c h^2$, $\theta_\textrm{MC}$, $\tau$, $A_s$, $n_s$),
where $\Omega_b$ and $\Omega_c$ are the current values of baryonic and cold dark matter density parameters, $h$ is the Hubble constant $H_0$ in units of
$100~\textrm{km}~\textrm{s}^{-1}~\textrm{Mpc}^{-1}$, $\theta_\textrm{MC}$ is the apparent size of the sound horizon at recombination defined
in the CAMB/COSMOMC program, $\tau$ is the reionization optical depth, and $A_s$ and $n_s$ are the amplitude and the spectral index of the primordial 
scalar-type energy density perturbation power spectrum.
In the tilted flat XCDM parameterization, we add one more free parameter, the equation of
state parameter ($w=p_X /\rho_X$, where $p_X$ and $\rho_X$ are the pressure and energy
density of the dark energy $X$-fluid). The $X$-fluid dark energy goes to the cosmological constant dark energy in the limit of $w=-1$.
In the tilted flat $\phi$CDM model where the scalar field potential energy density is
given by $V(\phi)=V_0 \phi^{-\alpha}$ \cite{PR1988}, we instead add the positive slope parameter $\alpha$ as a free parameter. The scalar field dark energy goes to the cosmological constant dark energy in the limit of $\alpha=0$.
In the untilted non-flat $\Lambda$CDM, XCDM, and $\phi$CDM models, the spectral index $n_s$ is replaced with the present value of the spatial curvature parameter $\Omega_k$.

The background evolution of the $\Lambda$CDM, XCDM, and $\phi$CDM dark energy models can be described by the evolution of the Hubble parameter. In the matter and dark energy dominated era, where the effect of radiation components can be ignored, the Hubble parameter at redshift $z$ is given by
\begin{equation}
   H^2 (a) = H_0^2 \left( \Omega_m a^{-3} + \Omega_k a^{-2} +\Omega_\Lambda  \right),
\end{equation}
and 
\begin{equation}
   H^2 (a) = H_0^2 \left( \Omega_m a^{-3} + \Omega_k a^{-2} +\Omega_X a^{-3(1+w)}  \right),
\end{equation}
for the $\Lambda$CDM and XCDM models, respectively. Here $a=1/(1+z)$ is the cosmic scale factor normalized to unity at present, and $\Omega_m=\Omega_b + \Omega_c$ and $\Omega_\Lambda$ ($\Omega_X$) are the present values of the matter and dark energy density parameters of the $\Lambda$CDM (XCDM) model, respectively. For the $\phi$CDM model, the Hubble parameter can be written as 
\begin{equation}
   H^2 (a) = \frac{H_0^2}{1-\frac{1}{6} (\phi^\prime)^2} \left[ \Omega_m a^{-3} + \Omega_k a^{-2} + \frac{V(\phi)}{3H_0^2} \right]
\end{equation}
with the equation of motion of the dark energy scalar field 
\begin{equation}
   \phi^{\prime\prime}+\left(3+\frac{\dot{H}}{H^2} \right) \phi^\prime + \frac{V_{,\phi}}{H^2}=0,
\end{equation} 
where $\phi^\prime \equiv d\phi / d\ln a$ and $V_{,\phi} =-V_0 \alpha \phi^{-\alpha-1}$.
Unlike in the $\Lambda$CDM and XCDM models, in the $\phi$CDM model we use the Hubble constant ($H_0$) as a free parameter
instead of $\theta_{\textrm{MC}}$ because the latter parameter is not suitable for use in some extreme situations where for larger values of $\alpha$ the 
scalar field dark energy density dominates in the early universe
(see Ref.\ \cite{ParkRatra2018c} for more details).

During the parameter exploration using the MCMC method, we set priors on 
parameters. We restrict the range of the Hubble constant to $0.2 \le h \le 1.0$, the reionization optical depth to $\tau \ge 0.005$, the dark energy equation of state parameter in the XCDM parameterization to $-3 \le w \le 0.2$, and the slope of the inverse power-law scalar field potential energy density index to $0 \le \alpha \le 10$ with flat priors. 
We also apply flat priors to other cosmological parameters, with sufficiently 
wide ranges so that they do not affect parameter estimation: e.g., $0.005 \le \Omega_b h^2 \le 0.1$, $0.001 \le \Omega_c h^2 \le 0.99$, $0.5 \le 100 \theta_\textrm{MC} \le 10$, $-0.5 \le \Omega_k \le 0.5$, and $0.8 \le n_s \le 1.2$, etc.

In our analysis, we consider the contribution of photons, massless and massive neutrinos by assuming the present CMB
temperature $T_0=2.7255$ K, the effective number of neutrino species $N_\textrm{eff}=3.04$, and a single massive neutrino
species with neutrino mass $0.06$ eV. 

The SPTpol data alone is not able to tightly constrain the reionization optical depth ($\tau$) as it is not so sensitive to $\tau$, and the $\tau$ parameter
is strongly correlated with $A_s$. When we constrain the parameters of each model using the SPTpol data alone
we use a Gaussian prior for $\tau$, adapted from the $\tau$ value of the corresponding model
constrained using the Planck 2015 CMB and the non-CMB data sets, and instead of $A_s$ and $\tau$ we 
use the combination $10^{9} A_s e^{-2\tau}$ as an alternative free parameter following the SPTpol team \cite{Henningetal2018}.
However, it should be emphasized that the resulting SPTpol parameter constraints strongly depend on the choice of the prior of $\tau$.

We use the converged MCMC chains to present mean values, their confidence limits, and likelihood distributions of the model
parameters. The convergence of the MCMC chains are checked with the Gelman and Rubin $R$ statistic using the COSMOMC getdist routine.

\section{Model Power Spectra}

Quantum-mechanical fluctuations during inflation generate the primordial 
energy density spatial inhomogeneity power spectra we use in our analyses here.

An initial epoch of non-slow-roll (tilted) spatially-flat inflation is used
to produce the primordial power spectrum in the spatially-flat models 
\cite{EI},
\begin{equation}
   P(k)=A_s \left(\frac{k}{k_0} \right)^{n_s},
\label{eq:Pk}
\end{equation}
where $k$ is wavenumber and the pivot wavenumber $k_0=0.05~\textrm{Mpc}^{-1}$.
This model and power spectrum expression remain valid in the slow-roll limit where $n_s = 1$ and they are untilted.
An initial epoch of slow-roll (untilted) non-flat inflation is used to produce
the power spectrum in the non-flat models 
\cite{Gott1982, closed, RatraPeebles1995, Ratra2017},
\begin{equation}
   P(q) \propto \frac{(q^2-4K)^2}{q(q^2-K)},
\label{eq:Pq}
\end{equation}
where $q=\sqrt{k^2 + K}$ is the non-flat space wavenumber and $K=-(H_0^2 / c^2) \Omega_k$ is the spatial curvature.
In the negative $\Omega_k$ closed model, the 
eigenvalue of the spatial Laplacian is 
$\propto -(q^2 - K)/K \equiv - {\bar k}^2/K$
and the normal modes are labeled by $q K^{-1/2}=3,4,5,\cdots$.  
In the non-flat models this $P(q)$ is normalized to $A_s$ at the 
$k_0$ pivot wavenumber. In the spatially-flat $K=0$ limit this $P(q)$
reduces to the untilted $n_s=1$ power spectrum. Unlike in the flat case, in the non-flat case it has not yet been possible to determine the power spectrum in a non-slow-roll inflation model.

On the other hand, the Planck non-flat model analyses 
\cite{PlanckCollaboration2016, PlanckCollaboration2018} do not make use of 
either of the above power spectra. They instead use
\begin{equation}
   P_{\rm Planck}(q) \propto \frac{(q^2-4K)^2}{q(q^2-K)} \left(\frac{\bar k}{k_0} \right)^{n_s-1}, 
\label{eq:PqP}
\end{equation}
where besides using the non-flat space wavenumber $q$, the wavenumber 
$\bar k$ is also used to define (and tilt) the non-flat model 
$P(q)$. The ${\bar k}^{n_s-1}$ tilt factor in $P_{\rm Planck}(q)$ is based 
on the assumption that tilt in a non-flat model works somewhat like it 
does in a flat model. This does not seem likely given that spatial 
curvature introduces an additional length scale in the non-flat case 
(i.e., in addition to the Hubble length). It is not known if the 
primordial power spectrum of Eq.~(\ref{eq:PqP}) is the consequence of quantum 
fluctuations during an epoch of inflation. However, this power spectrum is 
physically consistent if $n_s = 1$ or if $K= 0$, when it reduces to the 
power spectrum of Eqs.~(\ref{eq:Pq}) or (\ref{eq:Pk}), both of which are 
consequences of quantum fluctuations during inflation.

\section{Observational Constraints}

\begin{figure*}[htbp]
\centering
\mbox{\includegraphics[width=150mm]{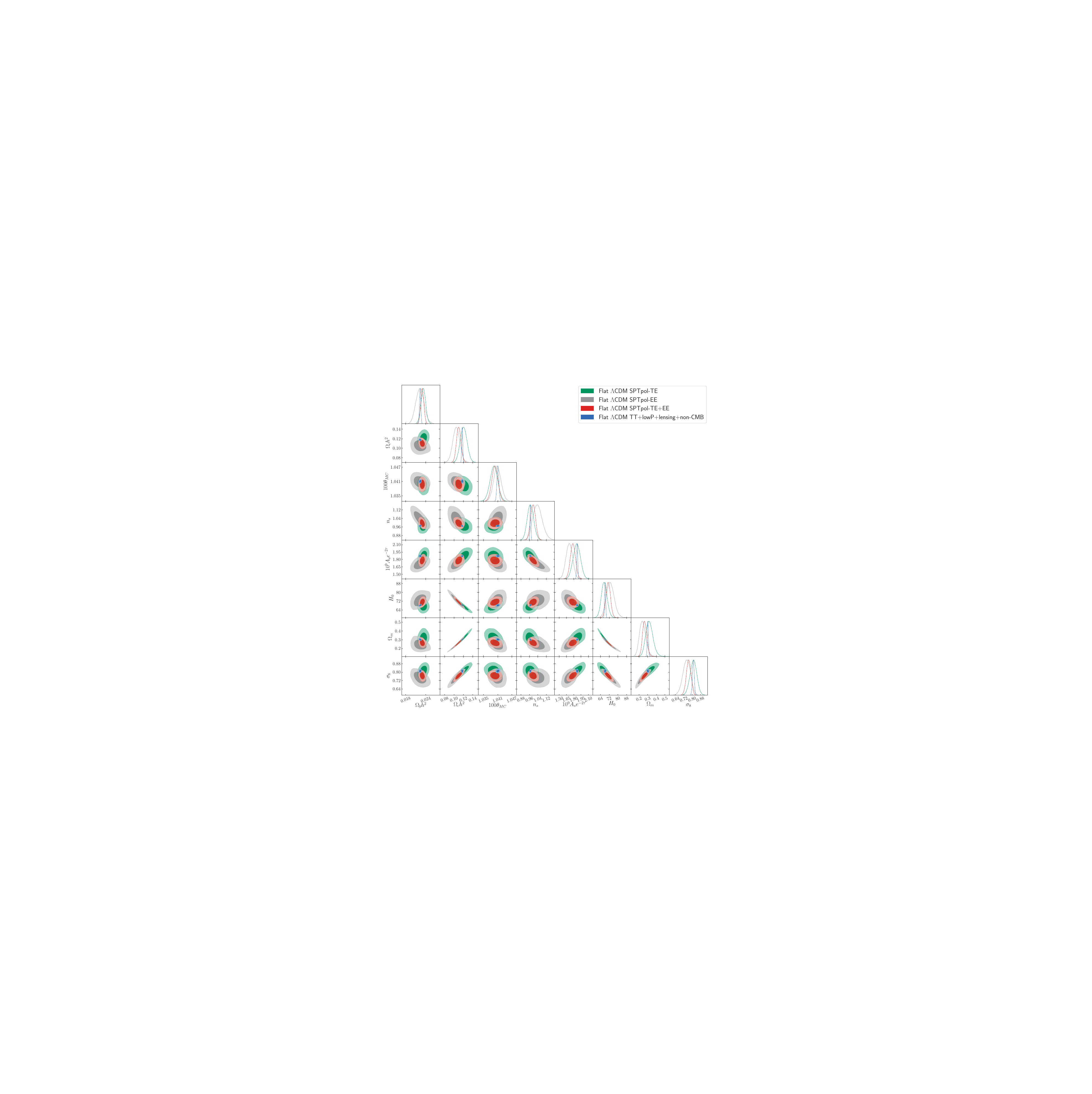}}
\caption{Likelihood distributions of the tilted flat $\Lambda$CDM model
         parameters constrained by using the SPTpol TE, EE, and TE+EE data.
         For comparison, results from the Planck 2015 data (TT+lowP+lensing) 
         together with non-CMB data sets (BAO, SN, $H(z)$, $f\sigma_8$) are 
         also shown. The red contours here are very similar to the 
         equivalent grey contours in Fig.\ 12 of Ref.\ \cite{Henningetal2018}. 
}
\label{fig:like_FL}
\end{figure*}

\begin{figure*}[htbp]
\centering
\mbox{\includegraphics[width=150mm]{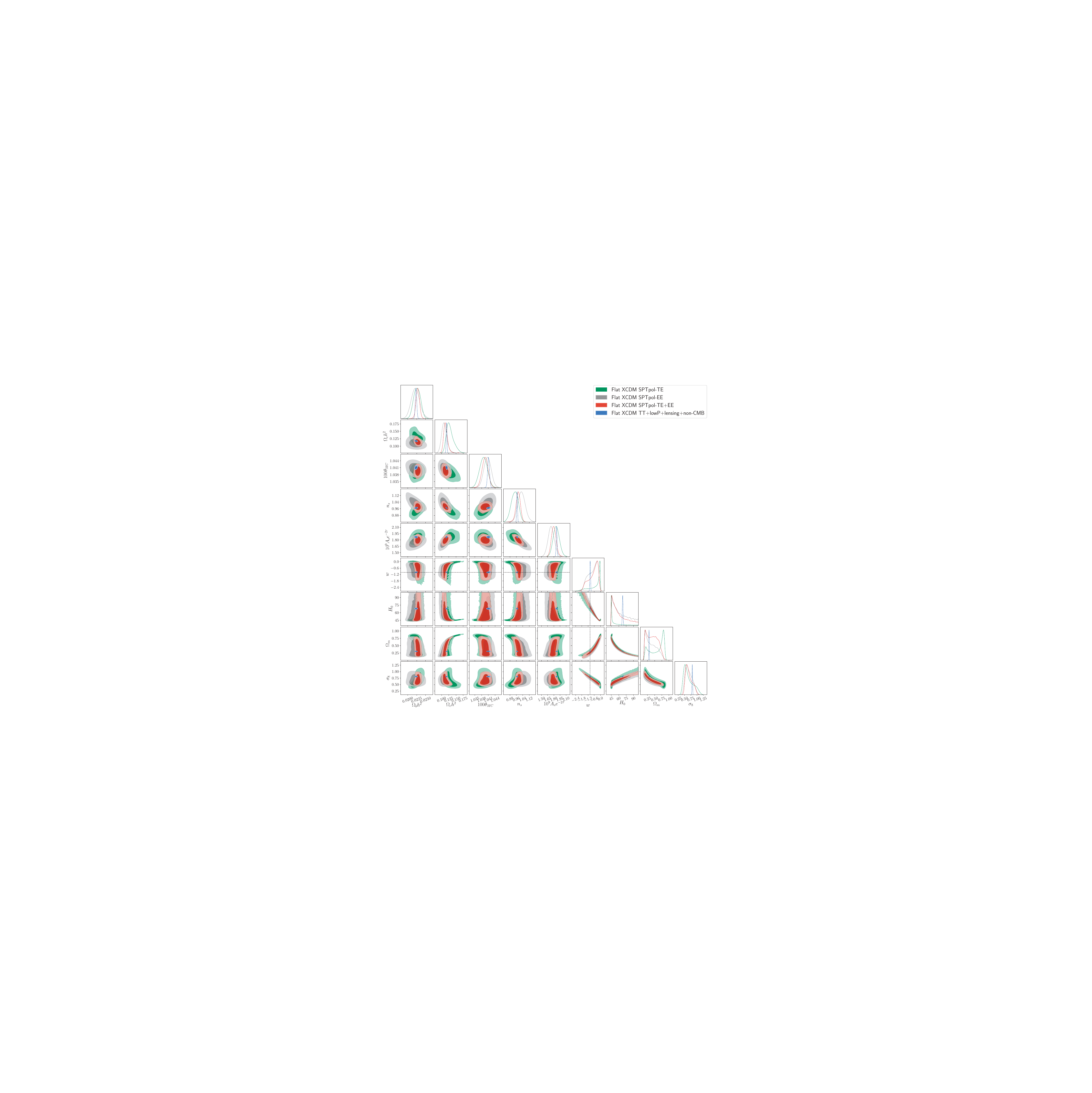}}
\caption{Likelihood distributions of the tilted flat XCDM model
         parameters constrained by using the SPTpol TE, EE, and TE+EE data.
         For comparison, results from the Planck 2015 data (TT+lowP+lensing) 
         together with non-CMB data sets (BAO, SN, $H(z)$, $f\sigma_8$) are 
         also shown. Dotted straight lines indicate $w=-1$.
}
\label{fig:like_FX}
\end{figure*}

\begin{figure*}[htbp]
\centering
\mbox{\includegraphics[width=150mm]{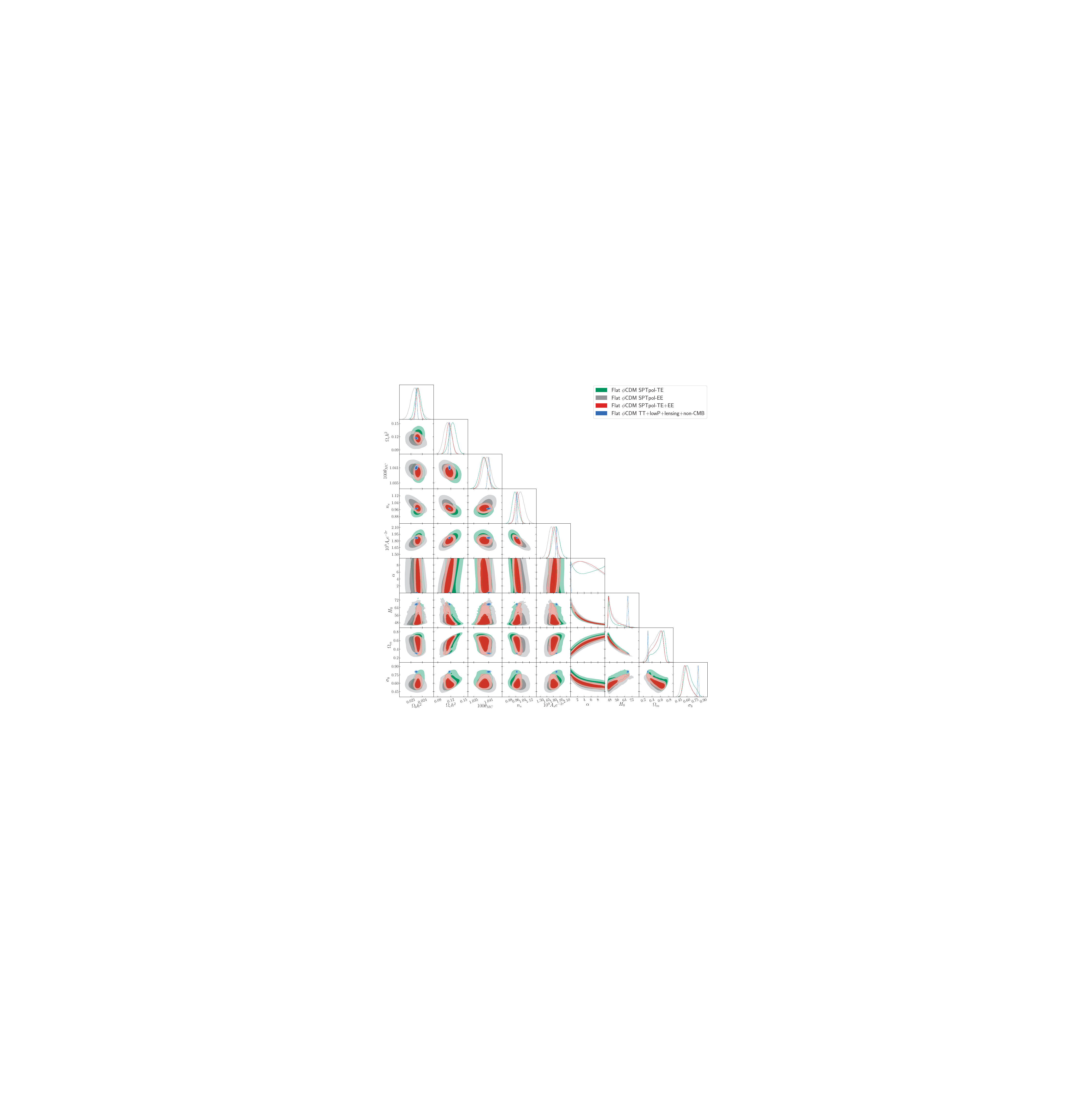}}
\caption{Likelihood distributions of the tilted flat $\phi$CDM model
         parameters constrained by using the SPTpol TE, EE, and TE+EE data.
         For comparison, results from the Planck 2015 data (TT+lowP+lensing) 
         together with non-CMB data sets (BAO, SN, $H(z)$, $f\sigma_8$) are 
         also shown. In the $\phi$CDM model, $H_0$ is used as an active free 
         parameter instead of $100\theta_\textrm{MC}$, but 
         $100\theta_\textrm{MC}$ is shown here above $H_0$
         for ease of comparison with other figures.
}
\label{fig:like_FQ}
\end{figure*}

The cosmological parameter constraints we describe in what follows were derived using the inflation power spectrum of Eq.~(\ref{eq:Pk}) in the spatially-flat models and Eq.~(\ref{eq:Pq}) in the non-flat models. In both cases we consider three different dark energy models, a time-independent cosmological constant and dynamical $X$-fluid and scalar field dark energy densities.

Figures \ref{fig:like_FL}--\ref{fig:like_FQ} show the likelihood distributions 
of model parameters of the tilted flat $\Lambda$CDM, XCDM, and $\phi$CDM models
derived by using the SPTpol-TE, SPTpol-EE, and SPTpol-TE+EE data. As 
mentioned above, we use the combination $10^{9} A_s e^{-2\tau}$ as a free 
parameter. In addition to the other four or five free cosmological parameters
that characterize these models, these plots also show constraints on three
derived cosmological parameters, namely, $H_0$, the current matter density 
parameter ($\Omega_m$), and the current amplitude of mass fluctuation 
at $8~h^{-1}\textrm{Mpc}$ scale ($\sigma_8$). In each figure, the result from 
the joint analysis of Planck CMB data (TT+lowP+lensing) and non-CMB data sets 
is shown for comparison. We note that our results for the SPTpol-TE+EE data are very similar to those of Ref.\ \cite{Henningetal2018}.

Table \ref{tab:para_flat} lists mean values and 68.3\% confidence ranges of 
cosmological parameters of the tilted flat $\Lambda$CDM, XCDM, and $\phi$CDM
models constrained using the SPTpol TE+EE, TE, and EE data sets.
We note that a different Gaussian prior of reionization optical depth that best-fits the Planck CMB and non-CMB data
has been applied for each dark energy model.

In the tilted flat XCDM and $\phi$CDM models, the dark energy parameters
($w$ and $\alpha$), $H_0$, and $\Omega_m$ are not tightly constrained by the 
SPTpol data alone. The SPTpol data (TE+EE and EE) prefer a larger value of $H_0$
in the $\Lambda$CDM model (as found in Ref.\ \cite{Henningetal2018}) and lower 
values of $H_0$ in both dynamical dark energy models, and particularly in the
$\phi$CDM model the estimated SPTpol Hubble constant is significantly lower than
the best-fit value obtained using the Planck CMB and non-CMB data. 
However, we note that in the XCDM model high values of Hubble constant near
the upper bound of the prior are allowed and there are strong degeneracies
between the Hubble constant and $w$, $\Omega_m$, and $\sigma_8$: 
$w$ and $\Omega_m$ have negative correlation with $H_0$ while $\sigma_8$
has positive correlation. In the tilted flat $\phi$CDM model, the SPTpol data 
alone do not constrain $\alpha$, allowing larger values of $\alpha$ exceeding 
the upper bound of the prior ($\alpha <10$). Larger values of $\alpha$ 
correspond to smaller values of the Hubble constant. The $\phi$CDM model
SPTpol Hubble constant has similar correlations with $\Omega_m$ and $\sigma_8$,
as in the XCDM parameterization case.

\begin{table*}
\caption{Tilted flat $\Lambda\textrm{CDM}$, XCDM, and $\phi$CDM model parameters constrained by using SPTpol TE+EE, TE, and EE data
         (mean and 68.3\% confidence limits).}
\begin{ruledtabular}
\begin{tabular}{lcccc}
\\[-1mm]                         & \multicolumn{4}{c}{Tilted flat $\Lambda$CDM ($\tau=0.066 \pm 0.012$ \cite{ParkRatra2018b})}              \\[+1mm]
\cline{2-5}\\[-1mm]
  Parameter                      & SPTpol TE+EE           & SPTpol TE              &  SPTpol EE           & TT+lowP+lensing+Non-CMB \\[+1mm]
 \hline \\[-1mm]
  $\Omega_b h^2$                 & $0.02295 \pm 0.00048$  & $0.02328 \pm 0.00071$  & $0.0223 \pm 0.0012$  & $0.02232 \pm 0.00019$   \\[+1mm]
  $\Omega_c h^2$                 & $0.1103 \pm 0.0048$    & $0.1208 \pm 0.0074$    & $0.1051 \pm 0.0076$  & $0.1177 \pm 0.0011$     \\[+1mm]
  $100\theta_\textrm{MC}$        & $1.0398 \pm 0.0013$    & $1.0394 \pm 0.0016$    & $1.0409 \pm 0.0016$  & $1.04108 \pm 0.00041$   \\[+1mm]
  $10^9 A_s e^{-2\tau}$          & $1.782 \pm 0.053$      & $1.863 \pm 0.075$      & $1.707 \pm 0.068$    & $1.871 \pm 0.011$       \\[+1mm]
  $n_s$                          & $0.995 \pm 0.024$      & $0.971 \pm 0.031$     & $1.038 \pm 0.045$    & $0.9692 \pm 0.0043$     \\[+1mm]
 \hline \\[-1mm]
  $H_0$ [km s$^{-1}$ Mpc$^{-1}$] & $71.1 \pm 2.1$         & $67.4 \pm 2.7$         & $73.1 \pm 3.7$       & $68.19 \pm 0.50$        \\[+1mm]
  $\Omega_m$                     & $0.266 \pm 0.025$      & $0.321 \pm 0.042$      & $0.243 \pm 0.038$    & $0.3025 \pm 0.0064$     \\[+1mm]
  $\sigma_8$                     & $0.767 \pm 0.023$      & $0.812 \pm 0.033$      & $0.747 \pm 0.038$    & $0.8117 \pm 0.0088$     \\[+1mm]
 \hline \\[-1mm]
                                 & \multicolumn{4}{c}{Tilted flat XCDM ($\tau=0.068 \pm 0.015$ \cite{ParkRatra2018b})}              \\[+1mm]
\cline{2-5}\\[-1mm]
  Parameter                      & SPTpol TE+EE           & SPTpol TE              &  SPTpol EE           & TT+lowP+lensing+Non-CMB \\[+1mm]
 \hline \\[-1mm]
  $\Omega_b h^2$                 & $0.02282 \pm 0.00049$  & $0.02261 \pm 0.00095$  & $0.0222 \pm 0.0012$  & $0.02233 \pm 0.00021$   \\[+1mm]
  $\Omega_c h^2$                 & $0.1146 \pm 0.0077$    & $0.130 \pm 0.013$      & $0.1097 \pm 0.0096$  & $0.1175 \pm 0.0014$     \\[+1mm]
  $100\theta_\textrm{MC}$        & $1.0395 \pm 0.0013$    & $1.0388 \pm 0.0017$    & $1.0406 \pm 0.0016$  & $1.04108 \pm 0.00042$   \\[+1mm]
  $10^9 A_s e^{-2\tau}$          & $1.800 \pm 0.059$      & $1.862 \pm 0.077$      & $1.738 \pm 0.077$    & $1.870 \pm 0.012$       \\[+1mm]
  $n_s$                          & $0.983 \pm 0.029$      & $0.940 \pm 0.044$      & $1.026 \pm 0.047$    & $0.9696 \pm 0.0051$     \\[+1mm]
  $w$                            & $-0.70 \pm 0.44$       & $-0.54 \pm 0.57$       & $-0.74 \pm 0.44$     & $-0.994 \pm 0.033$     \\[+1mm]
 \hline \\[-1mm]
  $H_0$ [km s$^{-1}$ Mpc$^{-1}$] & $62 \pm 15$            & $56 \pm 15$            & $64 \pm 16$          & $68.06 \pm 0.77$        \\[+1mm]
  $\Omega_m$                     & $0.42 \pm 0.18$        & $0.57 \pm 0.24$        & $0.38 \pm 0.19$      & $0.3034 \pm 0.0073$     \\[+1mm]
  $\sigma_8$                     & $0.68 \pm 0.13$        & $0.66 \pm 0.19$        & $0.68 \pm 0.14$      & $0.810 \pm 0.011$       \\[+1mm]
 \hline \\[-1mm]
                                 & \multicolumn{4}{c}{Tilted flat $\phi$CDM ($\tau=0.074 \pm 0.014$ \cite{ParkRatra2018c})}              \\[+1mm]
\cline{2-5}\\[-1mm]
  Parameter                      & SPTpol TE+EE           & SPTpol TE              &  SPTpol EE           & TT+lowP+lensing+Non-CMB \\[+1mm]
 \hline \\[-1mm]
  $\Omega_b h^2$                 & $0.02280 \pm 0.00048$  & $0.02282 \pm 0.00075$  & $0.0222 \pm 0.0012$  & $0.02238 \pm 0.00020$   \\[+1mm]
  $\Omega_c h^2$                 & $0.1164 \pm 0.0061$    & $0.1246 \pm 0.0082$    & $0.1124 \pm 0.0086$  & $0.1168 \pm 0.0013$     \\[+1mm]
  $H_0$ [km s$^{-1}$ Mpc$^{-1}$] & $52.0 \pm 5.6$         & $51.3 \pm 5.9$         & $52.4 \pm 6.2$       & $67.63 \pm 0.62$        \\[+1mm]
  $10^9 A_s e^{-2\tau}$          & $1.810 \pm 0.055$      & $1.859 \pm 0.076$      & $1.756 \pm 0.072$    & $1.867 \pm 0.011$       \\[+1mm]
  $n_s$                          & $0.977 \pm 0.026$      & $0.952 \pm 0.034$      & $1.017 \pm 0.044$    & $0.9715 \pm 0.0045$     \\[+1mm]
  $\alpha$ [95.4\% C.L.]         & $ < 9.6$               & $ < 9.8$               & $ < 9.7$             & $<0.22$                 \\[+1mm]
 \hline \\[-1mm]
  $100\theta_\textrm{MC}$        & $1.0392 \pm 0.0013$    & $1.0390 \pm 0.0016$    & $1.0402 \pm 0.0016$  & $1.04101 \pm 0.00042$   \\[+1mm]
  $\Omega_m$                     & $0.54 \pm 0.11$        & $0.58 \pm 0.12$        & $0.51 \pm 0.12$      & $0.3059 \pm 0.0068$     \\[+1mm]
  $\sigma_8$                     & $0.599 \pm 0.061$      & $0.636 \pm 0.078$      & $0.581 \pm 0.068$    & $0.8055 \pm 0.0098$     \\[+0mm]
\end{tabular}
\\[+1mm]
\begin{flushleft}
Note: Parameter constraints for Planck 2015 TT+lowP+lensing and non-CMB (SN, BAO, $H(z)$, $f\sigma_8$) data sets are from Ref.\ \cite{ParkRatra2018b} for the $\Lambda$CDM and XCDM models and Ref.\ \cite{ParkRatra2018c} for the $\phi$CDM model. For the SPTpol analyses, a different Gaussian prior for $\tau$ (indicated in the subheadings) has been used for each cosmological model (see main text for discussion and details).
\end{flushleft}
\end{ruledtabular}
\label{tab:para_flat}
\end{table*}

\begin{figure*}[htbp]
\centering
\mbox{\includegraphics[width=150mm]{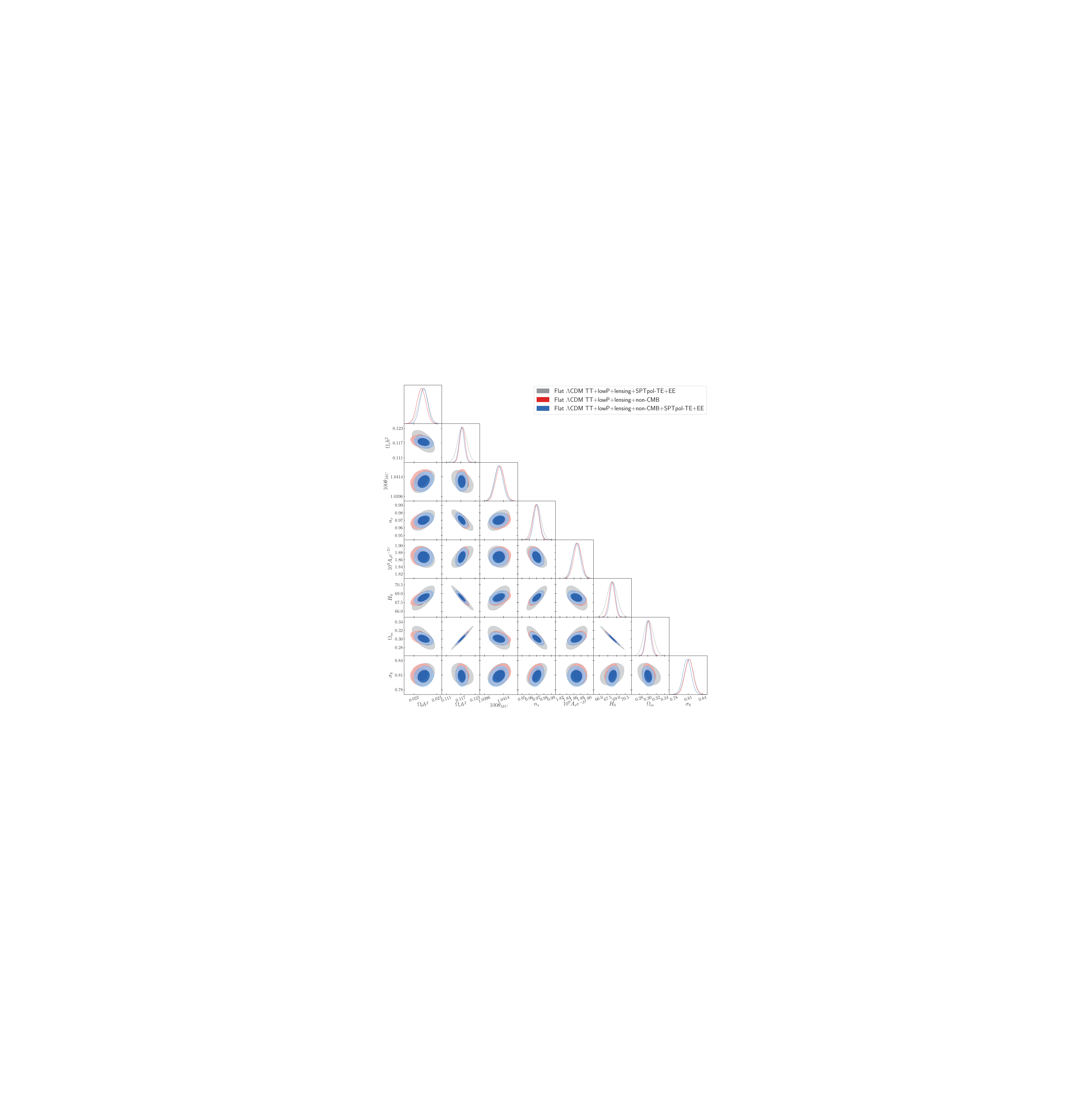}}
\caption{Likelihood distributions of the tilted flat $\Lambda$CDM model
         parameters constrained by using the Planck 2015 TT+lowP+lensing data 
         in conjunction with SPTpol TE+EE and non-CMB data.
}
\label{fig:like_FL_P2015}
\end{figure*}

\begin{figure*}[htbp]
\centering
\mbox{\includegraphics[width=150mm]{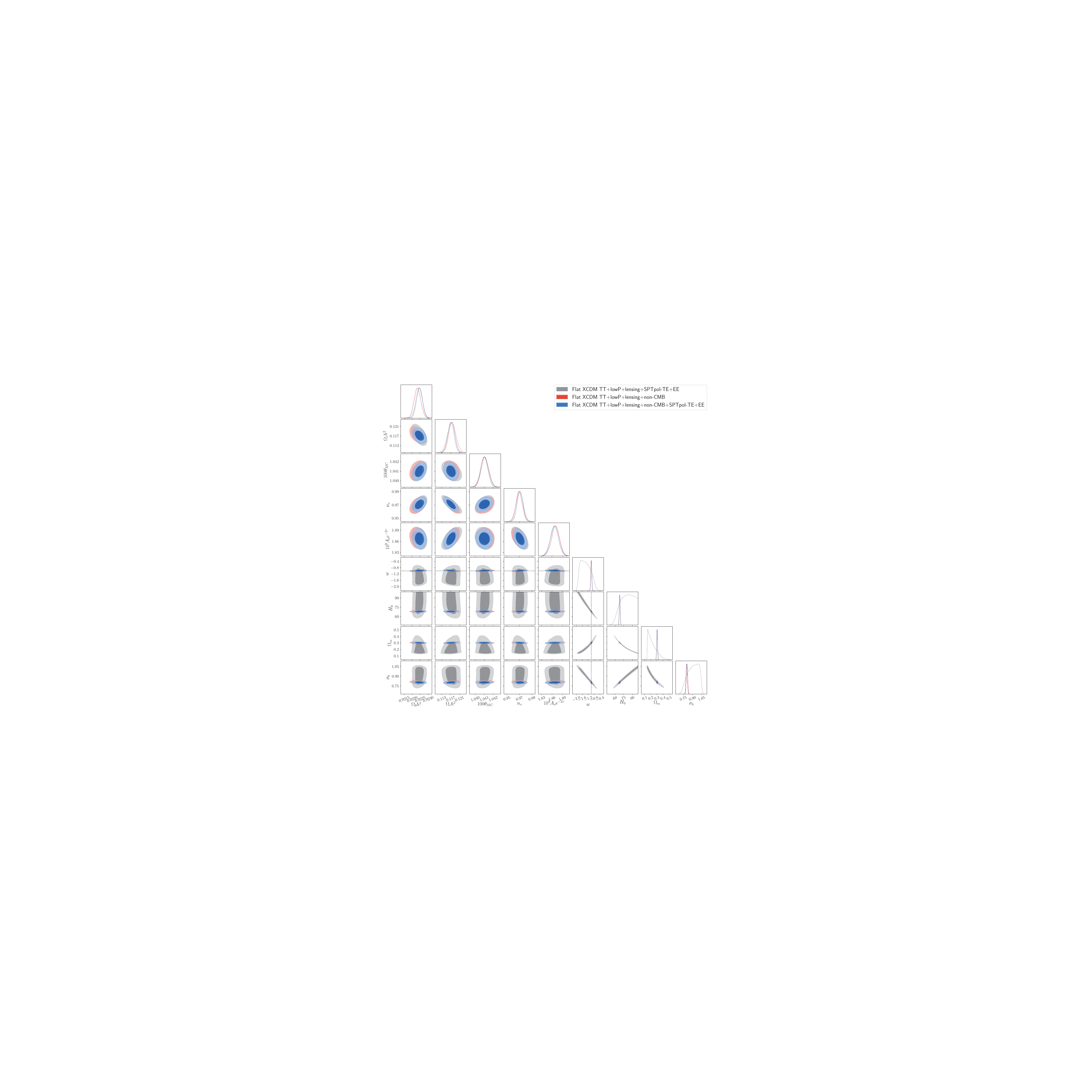}}
\caption{Likelihood distributions of the tilted flat XCDM model
         parameters constrained by using the Planck 2015 TT+lowP+lensing data 
         in conjunction with SPTpol TE+EE and non-CMB data. Dotted 
         straight lines indicate $w=-1$.
}
\label{fig:like_FX_P2015}
\end{figure*}

Figures \ref{fig:like_FL_P2015}--\ref{fig:like_FQ_P2015} show the likelihood
distributions of the tilted flat $\Lambda$CDM, XCDM, and $\phi$CDM model 
parameters constrained using different combinations of Planck 2015 CMB 
(TT+lowP+lensing), SPTpol, and non-CMB data. Here we do not apply a 
Gaussian prior on $\tau$ since the Planck CMB data provide a tight 
constraint on this parameter, but the combined parameter $10^9 A_s e^{-2\tau}$
is still used for comparison with the previous figures.
We see that the combination of Planck and SPTpol CMB data
(TT+lowP+lensing+SPTpol-TE+EE) is unable to place tight constraints 
on the dark energy parameters ($w$ and $\alpha$), $H_0$, $\Omega_m$, and 
$\sigma_8$ in the XCDM and $\phi$CDM dynamical dark energy models.

From our previous analysis of the CMB and non-CMB data 
\cite{ParkRatra2018b,ParkRatra2018c}, we note that, in conjunction with 
Planck data, BAO data are the most powerful in constraining these parameters.
Without the Planck CMB data sets, however, the BAO data alone (or combined with other non-CMB data) cannot give tight constraints on model parameters. Let us compare the case of SPTpol TE+EE data alone and the case of the SN+BAO+$H(z)$ data combination whose results are shown in Ref.\ \cite{ParkRatra2018d}. Both in the flat and the nonflat dark energy models, the SPTpol TE+EE data better constrains $\Omega_b h^2$, $\Omega_c h^2$, and $\Omega_k$ (for nonflat models) while the SN+BAO+$H(z)$ is better for the Hubble constant ($H_0$) and dark energy parameters $w$ and $\alpha$ (for XCDM and $\phi$CDM models) (see Table 2 of Ref.\ \cite{ParkRatra2018d} and Tables II and V of this paper). Thus, we cannot say that the statistical weight of the SPTpol data is lower than that of the BAO data.

Comparing the cases of TT+lowP+lensing+non-CMB and 
TT+lowP+lensing+non-CMB+SPTpol-TE+EE, we see that the non-CMB data are 
significantly better at helping constrain cosmological model parameters than 
are the SPTpol-TE+EE data, particularly in XCDM and $\phi$CDM models.
A summary of cosmological parameter values estimated from the three
different combination of data sets is given in Table \ref{tab:para_flat_planck}.
From this table and the figures, we see that, in the case of the tilted flat
dark energy models, the main effects of adding the SPTpol TE+EE data to the
TT+lowP+lensing+non-CMB data are a very slight tightening of the constraints
on $\Omega_b h^2$ and $\theta_\textrm{MC}$.

\begin{figure*}[htbp]
\centering
\mbox{\includegraphics[width=150mm]{fig6.pdf}}
\caption{Likelihood distributions of the tilted flat $\phi$CDM model
         parameters constrained by using the Planck 2015 TT+lowP+lensing data 
         in conjunction with SPTpol TE+EE and non-CMB data.
}
\label{fig:like_FQ_P2015}
\end{figure*}

We now examine the consistency between the tilted flat $\Lambda$CDM, XCDM, 
and $\phi$CDM 
model constraints determined from the SPTpol data and those determined from 
the Planck CMB and the non-CMB data sets. Following Ref.\ \cite{PlanckCollaboration2018} we quantify the consistency between the two sets of constraints by 
using the statistics of $\chi^2$'s for the parameter differences,
\begin{equation}
   \chi_p^2 = \Delta \mathbf{p}^T \mathbf{C}_p^{-1} \Delta \mathbf{p},
\end{equation}
where $\mathbf{C}_p$ is the covariance matrix of cosmological parameters
constrained by the SPTpol data alone and $\Delta \mathbf{p}$ denotes the
difference between the mean model parameters estimated from the SPTpol data 
and the best-fit ones estimated from the Planck 2015 CMB  and non-CMB data.
Following the Planck 2018 team's analysis \cite{PlanckCollaboration2018},
the small errors of parameters determined from the Planck CMB and non-CMB 
data are neglected. 

From the $\chi^2$ distribution with $k$ degrees of freedom, 
we compute the probability to exceed (PTE),
\begin{equation}
   \textrm{PTE}=1-P(k/2,\chi_p^2 /2)
\end{equation}
where $P(a,x)\equiv \gamma(a,x)/\Gamma(a)$ with the lower incomplete gamma 
function $\gamma(a,x)= \int_0^x e^{-t} t^{a-1} dt$ and the ordinary gamma 
function $\Gamma(a)=\gamma(a,\infty)$. The probability for exceeding the 
computed $\chi_p^2$ is estimated from the $\chi^2$ distribution with a given 
number of degrees of freedom ($k$), which is 5 for the tilted flat 
$\Lambda$CDM model and 6 for the tilted flat XCDM and $\phi$CDM models.
The results for the tilted flat $\Lambda$CDM, XCDM, and $\phi$CDM models
are summarized in the last two columns of Table \ref{tab:chi2_flat}. 
Here $N_b$ is the number of 
band powers of each SPTpol spectrum data set, $\chi_\textrm{min}^2$ denotes 
the minimum $\chi^2$ value for the particular data and model, and 
$N_\sigma=(\chi_\textrm{min}^2-N_\textrm{dof})/\sqrt{2 N_\textrm{dof}}$ 
indicates the deviation of $\chi_\textrm{min}^2$ from the expected value 
$\left<\chi_\textrm{min}^2 \right>=N_\textrm{dof}$. The number of degrees of 
freedom $N_\textrm{dof}$ is given by the number of data band powers minus 
the number of cosmological parameters (five for the $\Lambda$CDM
and six for the XCDM and $\phi$CDM models) and three nuisance parameters:  
$N_\textrm{dof}=N_b-8$ for the $\Lambda$CDM model and $N_\textrm{dof}=N_b-9$
for the XCDM and $\phi$CDM models (see Ref.\ \cite{PlanckCollaboration2018} for 
a detailed description).

The $\chi_\textrm{min}^2$ for the best-fit Planck 2018 cosmology or
the best-fit $\Lambda$CDM, XCDM, and $\phi$CDM models 
constrained by using the Planck 2015 CMB and non-CMB data is the 
minimum $\chi^2$ value and has been obtained by varying the SPTpol 
data-related nuisance and foreground parameters while fixing the best-fit 
cosmological parameters of the corresponding model. On the other hand,
the $\chi_\textrm{min}^2$ for the SPTpol data has been obtained by
varying all the cosmological, nuisance, and foreground parameters. Powell's minimization method (implemented in the COSMOMC program) has been used to find the best-fit model and the minimum value of the $\chi^2$.  
$\chi_\textrm{min}^2$ and $N_\sigma$ for the fits of the best-fit Planck
2018 tilted flat $\Lambda$CDM model and the best-fit TT+lowP+lensing+non-CMB tilted flat 
models to the SPTpol data are shown in the third and fourth columns of Table  
\ref{tab:chi2_flat}, while the corresponding numbers for the fits of the 
best-fit SPTpol tilted flat models to the SPTpol data are listed in the fifth
and sixth columns of this table.

\begin{table*}
\caption{Tilted flat $\Lambda\textrm{CDM}$, XCDM, and $\phi$CDM model parameters constrained by using Planck 2015, SPTpol TE+EE, and non-CMB data
         (mean and 68.3\% confidence limits).}
\begin{ruledtabular}
\begin{tabular}{lccc}
\\[-1mm]                         & \multicolumn{3}{c}{Tilted flat $\Lambda$CDM}              \\[+1mm]
\cline{2-4}\\[-1mm]
  Parameter                      & TT+lowP+lensing+SPTpol      & TT+lowP+lensing+non-CMB    & TT+lowP+lensing+non-CMB+SPTpol \\[+1mm]
 \hline \\[-1mm]
  $\Omega_b h^2$                 & $0.02241 \pm 0.00020$  &  $0.02232 \pm 0.00019$   &  $ 0.02243 \pm 0.00018$ \\[+1mm]
  $\Omega_c h^2$                 & $0.1176 \pm 0.0019$    &  $0.1177 \pm 0.0011$     &  $ 0.1174 \pm 0.0011$ \\[+1mm]
  $100\theta_\textrm{MC}$        & $1.04097 \pm 0.00043$  &  $1.04108 \pm 0.00041$   &  $ 1.04096 \pm 0.00038$ \\[+1mm]
  $\tau$                         & $0.067 \pm 0.016$      &  $0.066 \pm 0.012$       &  $ 0.064 \pm 0.012$ \\[+1mm]
  $\ln(10^{10} A_s)$             & $3.062 \pm 0.029$      &  $3.061 \pm 0.023$       &  $ 3.054 \pm 0.023$ \\[+1mm]
  $n_s$                          & $0.9702 \pm 0.0056$    &  $0.9692 \pm 0.0043$     &  $ 0.9703 \pm 0.0041$ \\[+1mm]
 \hline \\[-1mm]
  $10^9 A_s e^{-2\tau}$          & $1.868 \pm 0.013$      &  $1.871 \pm 0.011$       &  $ 1.867 \pm 0.011$ \\[+1mm]
  $H_0$ [km s$^{-1}$ Mpc$^{-1}$] & $68.24 \pm 0.85$       &  $68.19 \pm 0.50$        &  $ 68.34 \pm 0.48$ \\[+1mm]
  $\Omega_m$                     & $0.302 \pm 0.011$      &  $0.3025 \pm 0.0064$     &  $ 0.3008 \pm 0.0062$ \\[+1mm]
  $\sigma_8$                     & $0.8117 \pm 0.0093$    &  $0.8117 \pm 0.0088$     &  $ 0.8076 \pm 0.0087$ \\[+1mm]
 \hline \\[-1mm]
                                 & \multicolumn{3}{c}{Tilted flat XCDM}                \\[+1mm]
\cline{2-4}\\[-1mm]
  Parameter                      & TT+lowP+lensing+SPTpol      & TT+lowP+lensing+non-CMB     & TT+lowP+lensing+non-CMB+SPTpol \\[+1mm]
 \hline \\[-1mm]
  $\Omega_b h^2$                 & $0.02243 \pm 0.00020$  &  $0.02233 \pm 0.00021$   & $ 0.02245 \pm 0.00019$  \\[+1mm]
  $\Omega_c h^2$                 & $0.1175 \pm 0.0019$    &  $0.1175 \pm 0.0014$     & $ 0.1171 \pm 0.0014$  \\[+1mm]
  $100\theta_\textrm{MC}$        & $1.04102 \pm 0.0043$   &  $1.04108 \pm 0.00042$   & $ 1.04099 \pm 0.00040$  \\[+1mm]
  $\tau$                         & $0.060 \pm 0.017$      &  $0.068 \pm 0.015$       & $ 0.066 \pm 0.015$ \\[+1mm]
  $\ln(10^{10} A_s)$             & $3.049 \pm 0.031$      &  $3.063 \pm 0.027$       & $ 3.059 \pm 0.027$ \\[+1mm]
  $n_s$                          & $0.9705 \pm 0.0056$    &  $0.9696 \pm 0.0051$     & $ 0.9711 \pm 0.0047$  \\[+1mm]
  $w$                            & $-1.37 \pm 0.32$       &  $-0.994 \pm 0.033$      & $ -0.989 \pm 0.032$  \\[+1mm]
 \hline \\[-1mm]
  $10^9 A_s e^{-2\tau}$          & $1.868 \pm 0.013$      &  $1.870 \pm 0.012$       & $ 1.867 \pm 0.011$  \\[+1mm]
  $H_0$ [km s$^{-1}$ Mpc$^{-1}$] & $81 \pm 11$            &  $68.06 \pm 0.77$        & $ 68.10 \pm 0.76$  \\[+1mm]
  $\Omega_m$                     & $0.226 \pm 0.067$      &  $0.3034 \pm 0.0073$     & $ 0.3024 \pm 0.0071$  \\[+1mm]
  $\sigma_8$                     & $0.911 \pm 0.085$      &  $0.810 \pm 0.011$       & $ 0.805 \pm 0.011$  \\[+1mm]
 \hline \\[-1mm]
                                 & \multicolumn{3}{c}{Tilted flat $\phi$CDM}              \\[+1mm]
\cline{2-4}\\[-1mm]
  Parameter                      & TT+lowP+lensing+SPTpol  & TT+lowP+lensing+non-CMB   & TT+lowP+lensing+non-CMB+SPTpol \\[+1mm]
 \hline \\[-1mm]
  $\Omega_b h^2$                 & $0.02237 \pm 0.00021$  &  $0.02238 \pm 0.00020$   & $0.02250 \pm 0.00018$  \\[+1mm]
  $\Omega_c h^2$                 & $0.1182 \pm 0.0019$    &  $0.1168 \pm 0.0013$     & $0.1165 \pm 0.0013$  \\[+1mm]
  $H_0$ [km s$^{-1}$ Mpc$^{-1}$] & $63.3 \pm 3.4$         &  $67.63 \pm 0.62$        & $67.76 \pm 0.62$  \\[+1mm]
  $\tau$                         & $0.072 \pm 0.016$      &  $0.074 \pm 0.014$       & $0.072 \pm 0.014$ \\[+1mm]
  $\ln(10^{10} A_s)$             & $3.074 \pm 0.030$      &  $3.074 \pm 0.025$       & $3.069 \pm 0.025$ \\[+1mm]
  $n_s$                          & $0.9690 \pm 0.0057$    &  $0.9715 \pm 0.0045$     & $0.9726 \pm 0.0045$  \\[+1mm]
  $\alpha$ [95.4\% C.L.]         & $ < 1.62$              &  $<0.22$                 & $<0.22$  \\[+1mm]
 \hline \\[-1mm]
  $10^9 A_s e^{-2\tau}$          & $1.872 \pm 0.012$      &  $1.867 \pm 0.011$       & $1.864 \pm 0.011$  \\[+1mm]
  $100\theta_\textrm{MC}$        & $1.04070 \pm 0.00044$  &  $1.04101 \pm 0.00042$   & $1.04090 \pm 0.00039$  \\[+1mm]
  $\Omega_m$                     & $0.355 \pm 0.042$      &  $0.3059 \pm 0.0068$     & $0.3043 \pm 0.0067$  \\[+1mm]
  $\sigma_8$                     & $0.770 \pm 0.030$      &  $0.8055 \pm 0.0098$     & $0.8016 \pm 0.0096$  \\[+0mm]
\end{tabular}
\\[+1mm]
\begin{flushleft}
Note: Parameter constraints for Planck 2015 TT+lowP+lensing and non-CMB (SN, BAO, $H(z)$, $f\sigma_8$) data sets are from Ref.\ \cite{ParkRatra2018b} for the $\Lambda$CDM and XCDM models and Ref.\ \cite{ParkRatra2018c} for the $\phi$CDM model.
\end{flushleft}
\end{ruledtabular}
\label{tab:para_flat_planck}
\end{table*}

Our results for the Planck 2018 best-fit flat $\Lambda$CDM model and the 
SPTpol best-fit flat $\Lambda$CDM model with the Gaussian prior on $\tau$ 
used in \cite{Henningetal2018} are 
similar to those presented in the Planck 2018 analysis (Table 3 of Ref.\ \cite{PlanckCollaboration2018}).
The values of $\chi_\textrm{min}^2$ and $N_\sigma$ for the Planck 2018 best-fit 
cosmology are higher than those for the best-fit $\Lambda$CDM model constrained with SPTpol data,
which suggests that the SPTpol data disfavor the $\Lambda$CDM cosmology that 
best-fits the Planck 2018 data.
The same holds for the best-fit tilted flat $\Lambda$CDM, XCDM, and $\phi$CDM models as shown
in Table \ref{tab:chi2_flat}.
We note that in all cases of SPTpol TE+EE $N_\sigma$ is larger than 2.2 so the best-fit models do not provide good fits
to the SPTpol TE+EE data. The XCDM model fits the SPTpol TE+EE, TE, and EE 
data sets better than does the $\Lambda$CDM model.
Here $\chi_\textrm{min}^2=62.36$ for the best-fit flat XCDM model favored by the SPTpol TE data
is at least 5 less than the values of the best-fit $\Lambda$CDM and $\phi$CDM models. 
The $\phi$CDM model with larger $\chi_\textrm{min}^2$ and $N_\sigma$
poorly fits the SPTpol data.

Comparing $\chi_p^2$ values in the tilted flat $\Lambda$CDM, XCDM, and 
$\phi$CDM models implies there is a discrepancy between the best-fit model 
favored
by the Planck CMB and non-CMB data and that favored by the SPTpol data alone.
For the SPTpol TE+EE data, the discrepancy in the XCDM case is smaller
than that in the $\Lambda$CDM model. The $\phi$CDM model has an even bigger 
discrepancy than the other two models,
with larger $\chi_p^2$ and smaller PTE value, \cite{Table36note}. 
However, there is no significant evidence of tension between the $\phi$CDM 
(or $\Lambda$CDM or XCDM) model constrained using the Planck CMB and
non-CMB data and that constrained using SPTpol data alone (as can be 
seen in the last column of Table \ref{tab:chi2_flat}).

\begin{table*}
\caption{Minimum $\chi^2$ values for the SPTpol TE+EE, TE, and EE spectra in the best-fit tilted flat Planck 2018 $\Lambda$CDM model, and in the best-fit tilted
         flat $\Lambda$CDM, XCDM, and $\phi$CDM models constrained using Planck 2015 TT+lowP+SN+BAO+$H(z)$+$f\sigma_8$ data \cite{ParkRatra2018b,ParkRatra2018c}. }
\begin{ruledtabular}
\begin{tabular}{lrcccccc}
                       &       & \multicolumn{2}{c}{Planck 2018 cosmology \cite{PlanckCollaboration2018}} &  \multicolumn{2}{c}{SPT ($\Lambda$CDM $\tau$ \cite{Henningetal2018})}  & &  \\[+1mm]
\cline{3-4}\cline{5-6}\\[-1mm]
    SPTpol spectrum    & $N_b$ & $\chi_{\textrm{min}}^2$  & $N_{\sigma}$   & $\chi_{\textrm{min}}^2$  & $N_{\sigma}$ & $\chi_p^2$  &  PTE   \\[+1mm]
\hline\\[-1mm]
    TE + EE            & 112   & 146.80                   & 2.97           & 137.27                   & 2.31         &    9.95     & 0.077       \\[+1mm]
    TE                 &  56   &  71.62                   & 2.41           &  68.21                   & 2.06         &    3.26     & 0.659       \\[+1mm]
    EE                 &  56   &  67.34                   & 1.97           &  60.88                   & 1.31         &    8.74     & 0.120       \\[+1mm]
\hline\\[-1mm]
                       &       & \multicolumn{2}{c}{Best-fit tilted flat $\Lambda$CDM}   &  \multicolumn{2}{c}{SPT ($\Lambda$CDM $\tau$ \cite{ParkRatra2018b})} &   &  \\[+1mm]
\cline{3-4}\cline{5-6}\\[-1mm]
    SPTpol spectrum    & $N_b$ & $\chi_{\textrm{min}}^2$  & $N_{\sigma}$  & $\chi_{\textrm{min}}^2$  & $N_{\sigma}$ & $\chi_p^2$   &  PTE   \\[+1mm]
\hline\\[-1mm]
    TE + EE            & 112   & 146.84                   & 2.97          & 136.92                   & 2.28          &    8.41     & 0.135       \\[+1mm]
    TE                 &  56   &  72.22                   & 2.47          &  67.94                   & 2.04          &    3.75     & 0.586       \\[+1mm]
    EE                 &  56   &  67.34                   & 1.97          &  60.64                   & 1.29          &    7.80     & 0.168       \\[+1mm]
\hline\\[-1mm]
                       &       & \multicolumn{2}{c}{Best-fit tilted flat XCDM}     &  \multicolumn{2}{c}{SPT (XCDM $\tau$ \cite{ParkRatra2018b})}   &             &  \\[+1mm]
\cline{3-4}\cline{5-6}\\[-1mm]
    SPTpol spectrum    & $N_b$ & $\chi_{\textrm{min}}^2$  & $N_{\sigma}$  & $\chi_{\textrm{min}}^2$  & $N_{\sigma}$  & $\chi_p^2$  &  PTE   \\[+1mm]
\hline\\[-1mm]
    TE + EE            & 112   & 146.69                   & 3.04          & 135.31                   & 2.25          &    6.75     & 0.345       \\[+1mm]
    TE                 &  56   &  72.17                   & 2.60          &  62.36                   & 1.58          &    4.08     & 0.666       \\[+1mm]
    EE                 &  56   &  67.27                   & 2.09          &  59.03                   & 1.24          &    6.01     & 0.422       \\[+1mm]
\hline\\[-1mm]
                       &       & \multicolumn{2}{c}{Best-fit tilted flat $\phi$CDM}   &  \multicolumn{2}{c}{SPT ($\phi$CDM $\tau$ \cite{ParkRatra2018c})}  &         &  \\[+1mm]
\cline{3-4}\cline{5-6}\\[-1mm]
    SPTpol spectrum    & $N_b$ & $\chi_{\textrm{min}}^2$  & $N_{\sigma}$  & $\chi_{\textrm{min}}^2$  & $N_{\sigma}$  & $\chi_p^2$  &  PTE ($H_0$)  \\[+1mm]
\hline\\[-1mm]
    TE + EE            & 112   & 146.59                   & 3.04          & 139.27                   & 2.53          &   21.85     & 0.001  \\[+1mm]
    TE                 &  56   &  72.14                   & 2.59          &  68.21                   & 2.19          &   15.68     & 0.016  \\[+1mm]
    EE                 &  56   &  67.29                   & 2.09          &  61.28                   & 1.47          &   17.50     & 0.008  \\[+1mm]
\cline{3-4}\cline{5-6}\\[-1mm]
    SPTpol spectrum    & $N_b$ & $\chi_{\textrm{min}}^2$  & $N_{\sigma}$  & $\chi_{\textrm{min}}^2$  & $N_{\sigma}$  & $\chi_p^2$  &  PTE ($\theta_\textrm{MC}$)  \\[+1mm]
\hline\\[-1mm]
    TE + EE            & 112   & 146.59                   & 3.04          & 139.27                   & 2.53          &   10.84     & 0.093  \\[+1mm]
    TE                 &  56   &  72.14                   & 2.59          &  68.21                   & 2.19          &    6.18     & 0.403  \\[+1mm]
    EE                 &  56   &  67.29                   & 2.09          &  61.28                   & 1.47          &    9.99     & 0.125  \\[+0mm]
\end{tabular}
\\[+1mm]
\begin{flushleft}
Note: We assume a different Gaussian prior of $\tau$ for each cosmological model. For the best-fit tilted flat Planck 2018 model, and the best-fit tilted flat $\Lambda$CDM, XCDM, and $\phi$CDM models constrained using Planck 2015 and non-CMB data, we apply $\tau=0.078\pm 0.019$ \cite{Henningetal2018}, $0.066 \pm 0.012$, $0.068 \pm 0.015$ \cite{ParkRatra2018b}, and $0.074 \pm 0.014$ \cite{ParkRatra2018c}, respectively.
\end{flushleft}
\end{ruledtabular}
\label{tab:chi2_flat}
\end{table*}

\begin{figure*}[htbp]
\centering
\mbox{\includegraphics[width=150mm]{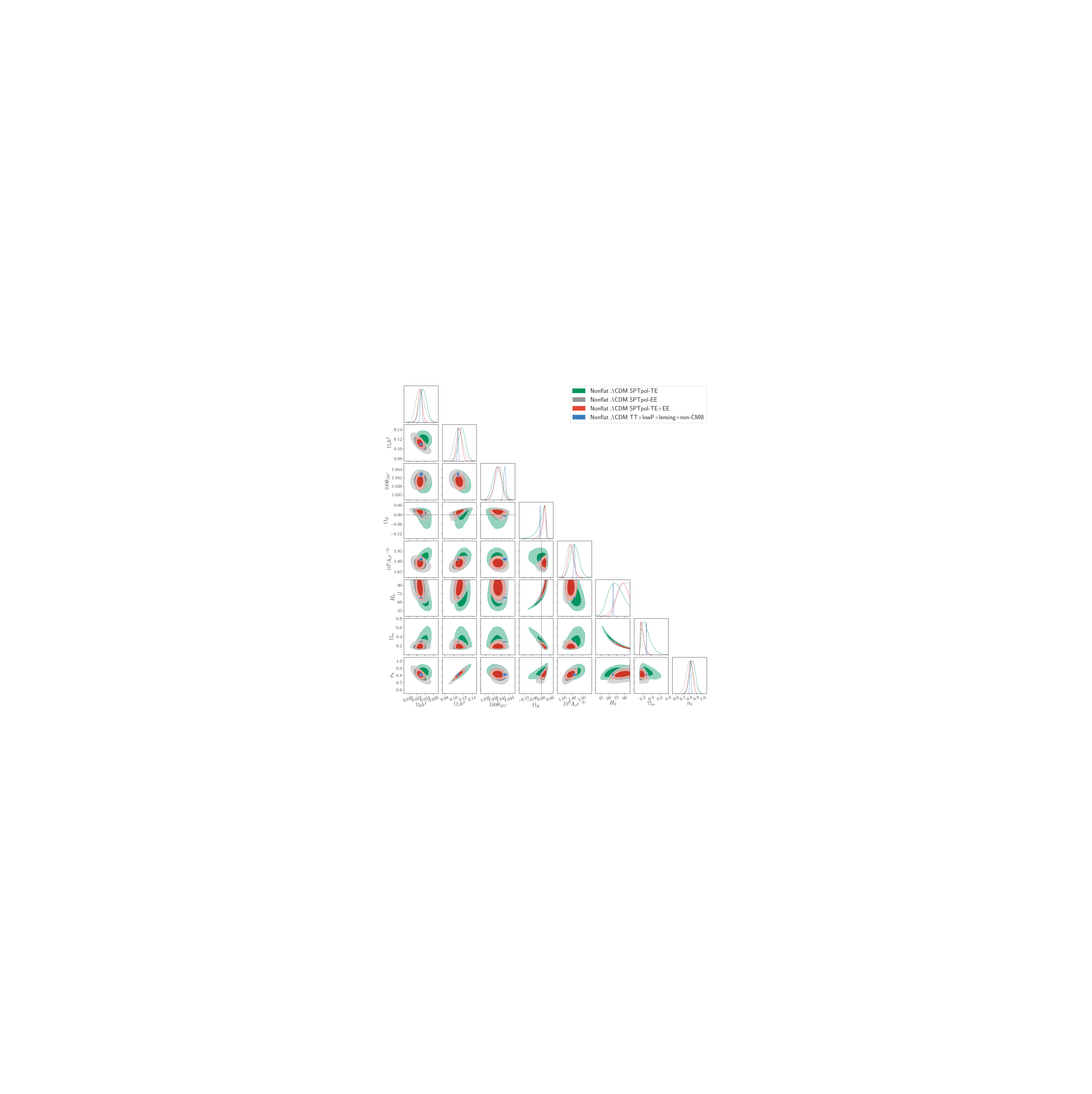}}
\caption{Likelihood distributions of the untilted non-flat $\Lambda$CDM model
         parameters constrained by using the SPTpol TE+EE, TE, and EE data 
         alone.
         For comparison, results from the Planck 2015 data (TT+lowP+lensing) 
         together with non-CMB data sets (BAO, SN, $H(z)$, $f\sigma_8$) are 
         also shown. Dotted straight lines indicate $\Omega_k$ = 0.
}
\label{fig:like_NL_SPT}
\end{figure*}

\begin{figure*}[htbp]
\centering
\mbox{\includegraphics[width=150mm]{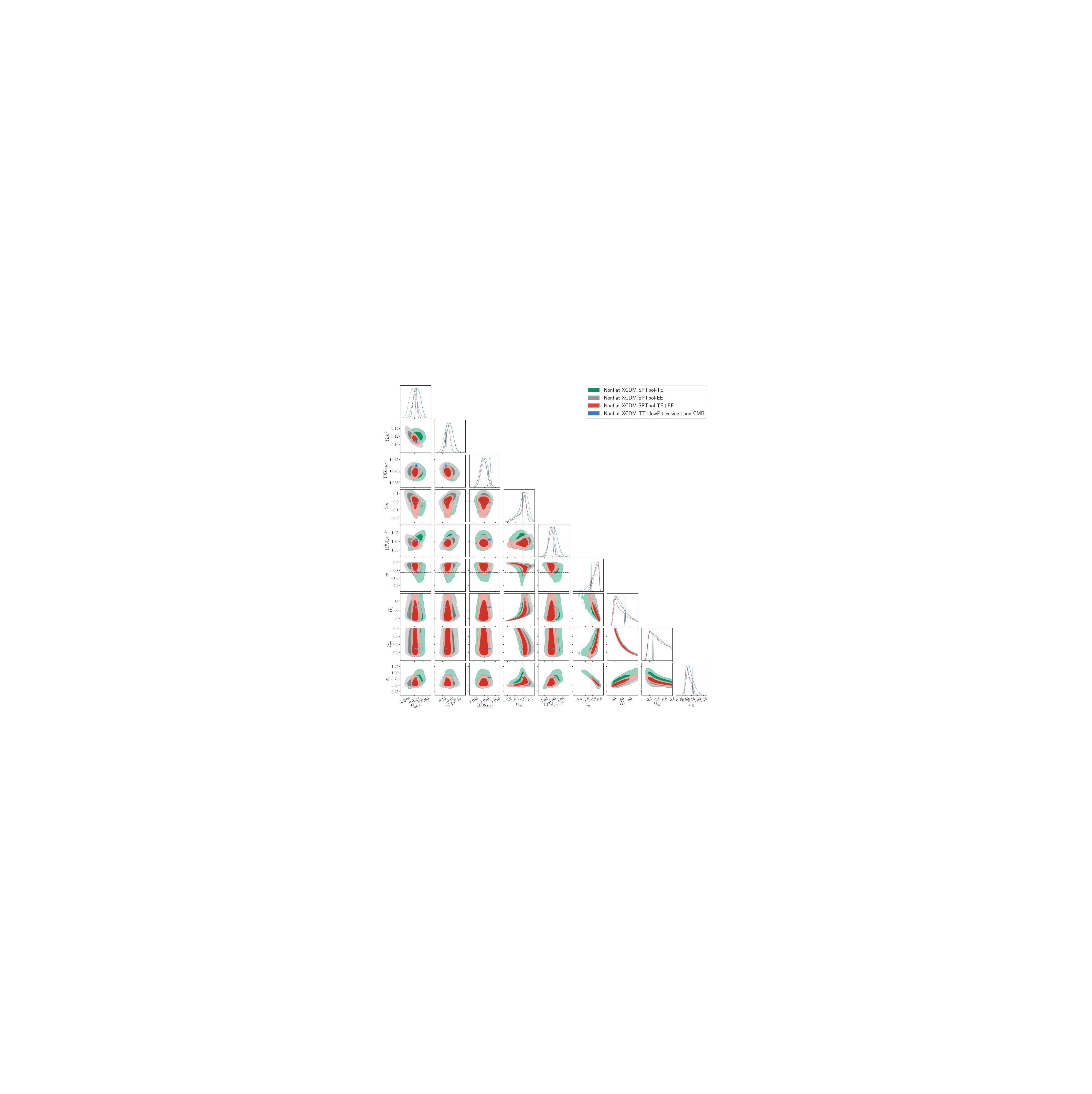}}
\caption{Likelihood distributions of the untilted non-flat XCDM model
         parameters constrained by using the SPTpol TE+EE, TE, and EE data 
         alone.
         For comparison, results from the Planck 2015 data (TT+lowP+lensing)
         together with non-CMB data sets (BAO, SN, $H(z)$, $f\sigma_8$) are 
         also shown. Dotted straight lines indicate $w=-1$ or $\Omega_k$ = 0.
}
\label{fig:like_NX_SPT}
\end{figure*}

\begin{figure*}[htbp]
\centering
\mbox{\includegraphics[width=150mm]{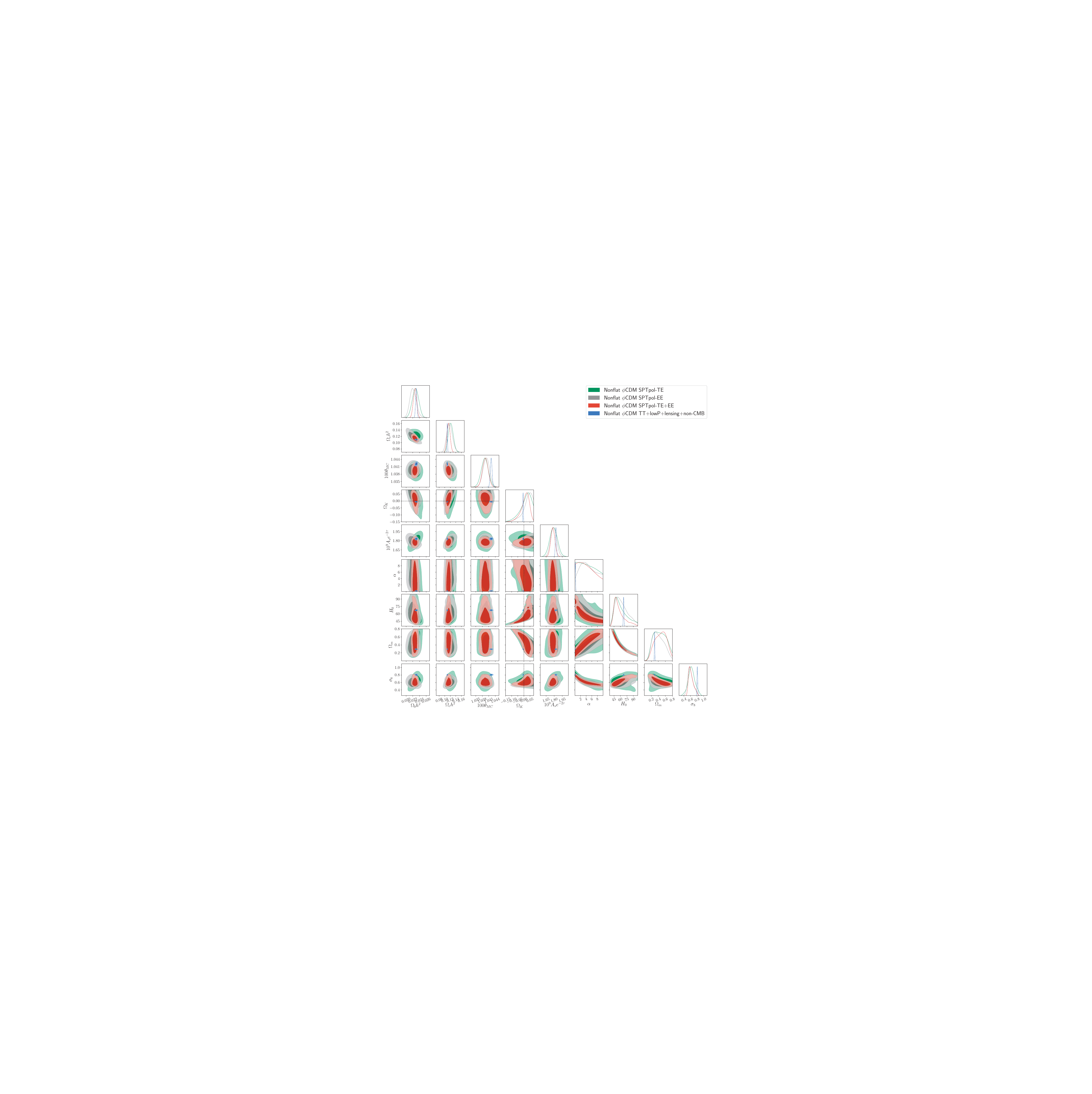}}
\caption{Likelihood distributions of the untilted non-flat $\phi$CDM model
         parameters constrained by using the SPTpol TE+EE, TE, and EE data 
         alone.
         For comparison, results from the Planck 2015 data (TT+lowP+lensing)
         together with non-CMB data sets (BAO, SN, $H(z)$, $f\sigma_8$) are 
         also shown. Dotted straight lines indicate $\Omega_k$ = 0.
}
\label{fig:like_NQ_SPT}
\end{figure*}

We now consider the untilted non-flat model constraints derived using the 
SPTpol data alone. Figures \ref{fig:like_NL_SPT}--\ref{fig:like_NQ_SPT} show 
the likelihood distributions of model parameters of the untilted non-flat 
$\Lambda$CDM, XCDM, and $\phi$CDM models that are favored by the SPTpol data 
sets. In each figure, the joint analysis results obtained by using the Planck 
2015 CMB and non-CMB data are also shown. 
Compared to the combination of the Planck CMB and non-CMB data,
the SPTpol data sets tend to prefer smaller values of $\theta_\textrm{MC}$.
As in the tilted flat models with SPTpol data, the observations
prefer a parameter space where the Hubble constant is anti-correlated
with the matter density $\Omega_m$ and has some degeneracy with the curvature
parameter $\Omega_k$.

The mean values and 68.3\% confidence ranges of the cosmological parameters 
of untilted non-flat dark energy models are summarized in Table 
\ref{tab:para_nonflat}. In the untilted non-flat $\Lambda$CDM model, 
the SPTpol data favor a larger Hubble constant value while the situation 
is the opposite in the XCDM and $\phi$CDM models. The SPTpol data by itself 
does not significantly constrain the spatial curvature parameter $\Omega_k$.

\begin{figure*}[htbp]
\centering
\mbox{\includegraphics[width=150mm]{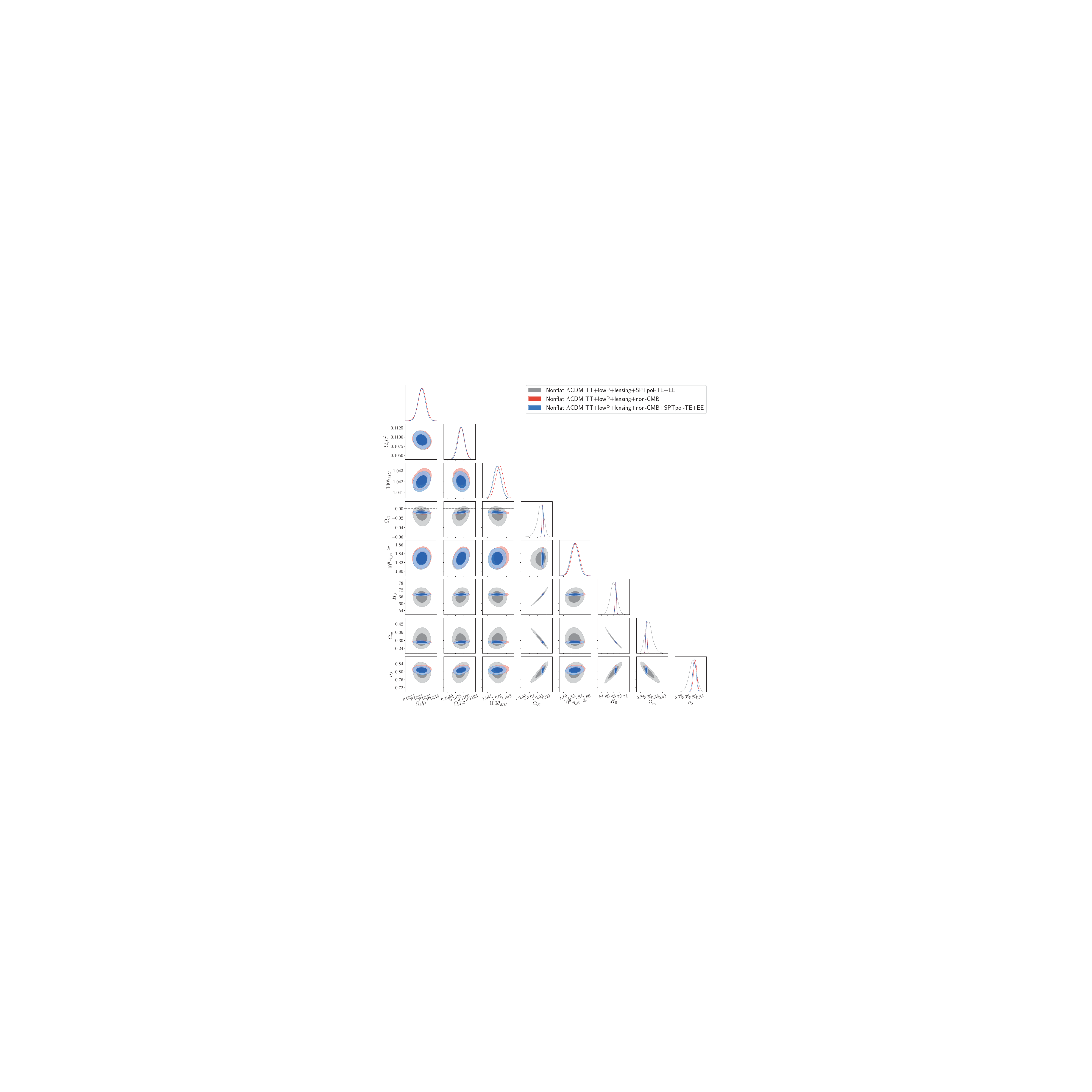}}
\caption{Likelihood distributions of the untilted non-flat $\Lambda$CDM model
         parameters constrained by using Planck 2015 TT+lowP+lensing data, 
         non-CMB  data, and SPTpol TE+EE data. Dotted straight lines indicate 
         $\Omega_k$ = 0.
}
\label{fig:like_NL_SPT_planck}
\end{figure*}

\begin{figure*}[htbp]
\centering
\mbox{\includegraphics[width=150mm]{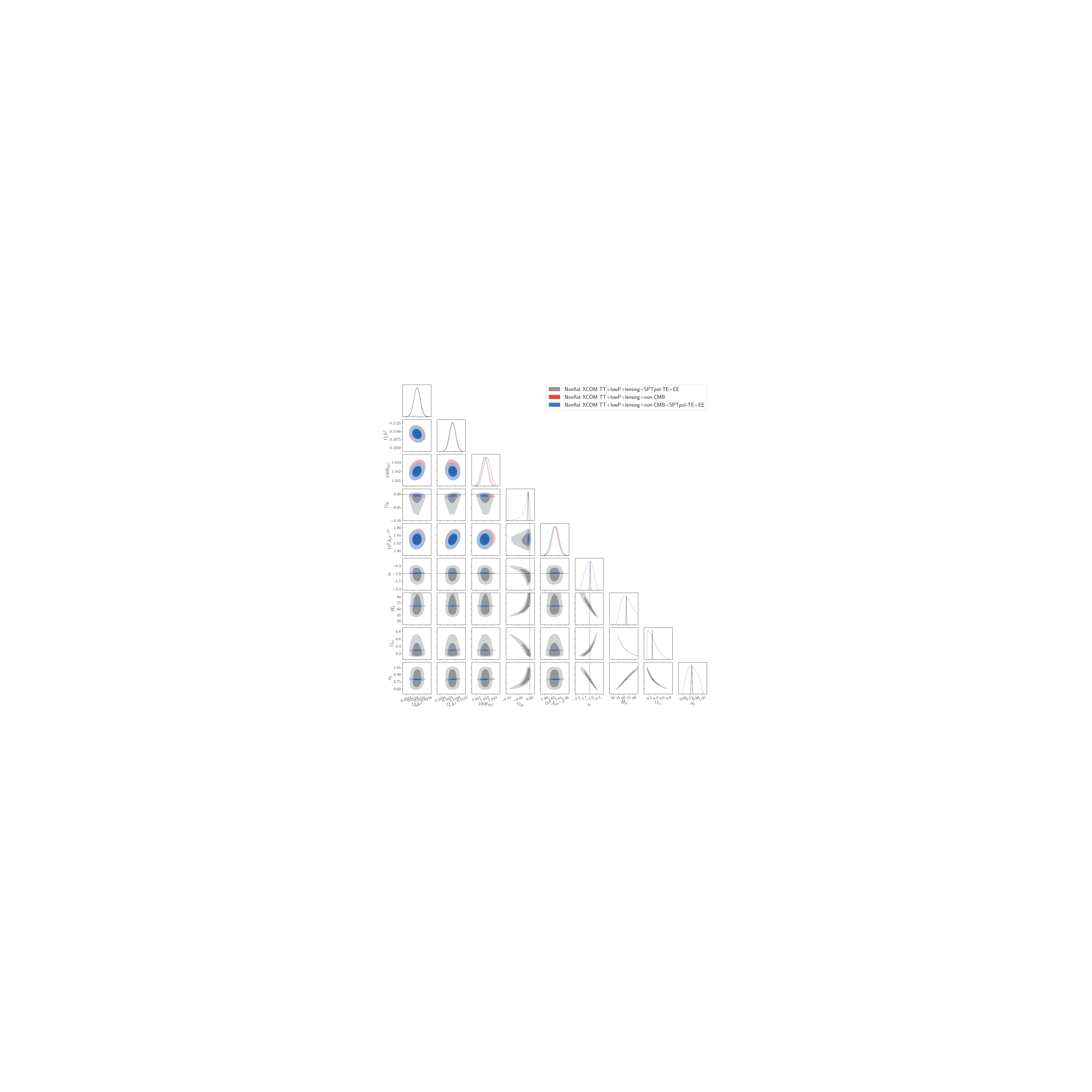}}
\caption{Likelihood distributions of the untilted non-flat XCDM model
         parameters constrained by using Planck 2015 TT+lowP+lensing data, 
         non-CMB data, and SPTpol TE+EE data. Dotted straight lines indicate $w=-1$ or 
         $\Omega_k$ = 0.
}
\label{fig:like_NX_SPT_planck}
\end{figure*}

\begin{figure*}[htbp]
\centering
\mbox{\includegraphics[width=150mm]{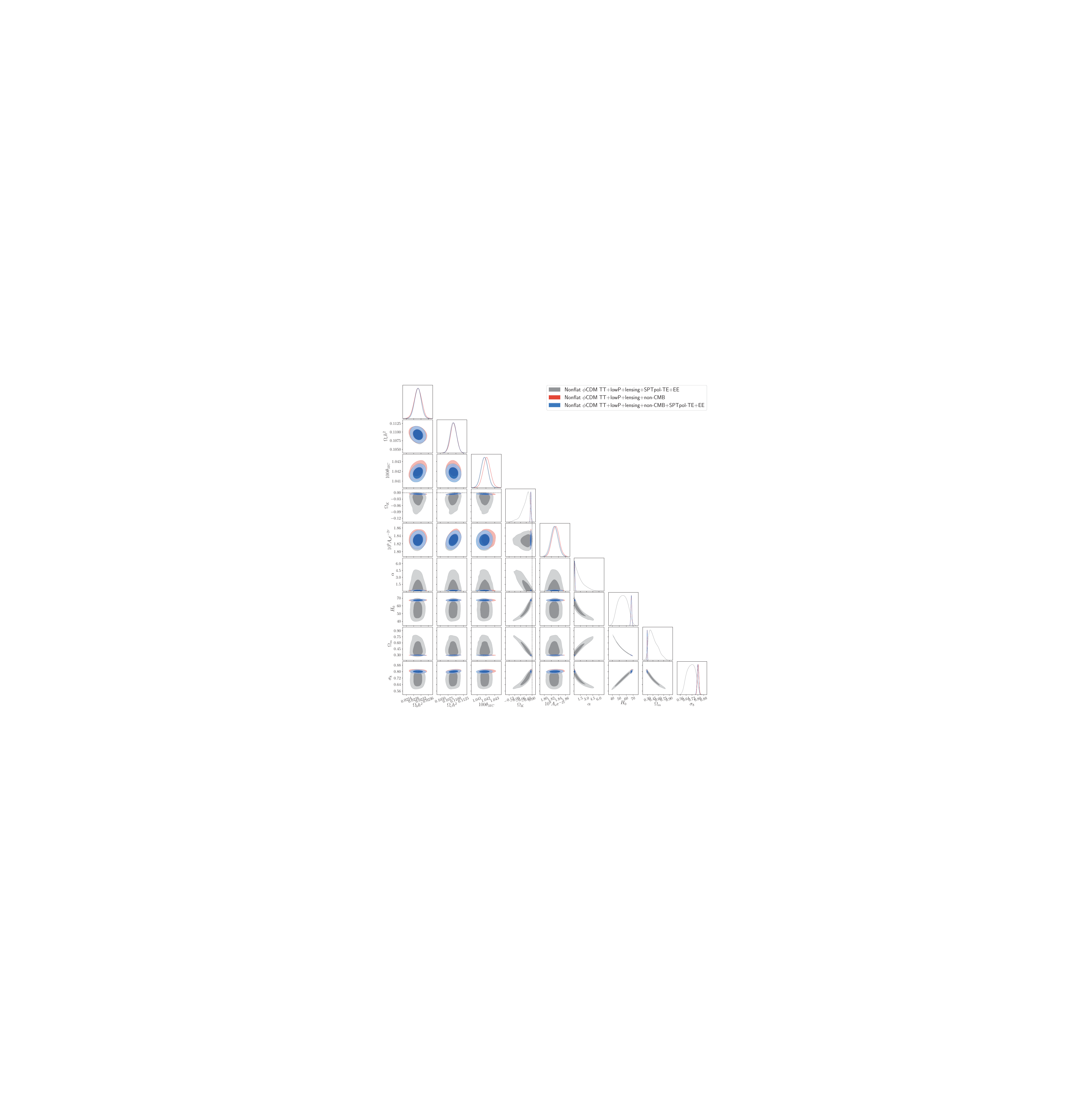}}
\caption{Likelihood distributions of the untilted non-flat $\phi$CDM model
         parameters constrained by using Planck TT+lowP+lensing data, 
         non-CMB data, and SPTpol TE+EE data. Dotted straight lines indicate 
         $\Omega_k$ = 0.
}
\label{fig:like_NQ_SPT_planck}
\end{figure*}

Figures \ref{fig:like_NL_SPT_planck}--\ref{fig:like_NQ_SPT_planck} show the
likelihood distributions of the untilted non-flat $\Lambda$CDM, XCDM, and
$\phi$CDM model parameters constrained using different combinations of Planck
2015 CMB (TT+lowP+lensing), SPTpol, and non-CMB data sets. Again, we do not
use a Gaussian prior on $\tau$ in these analyses since the Planck CMB data
provide a tight constraint on this parameter. We see that the combination
of Planck and SPTpol data (TT+lowP+lensing+SPTpol-TE+EE) is unable to place
tight constraints on the dark energy parameters ($w$ and $\alpha$), $H_0$,
$\Omega_k$, $\Omega_m$, and $\sigma_8$ in all three models. The degeneracies
between parameters seen in the constraints from the SPTpol data alone are also 
present here. Comparing the cases of TT+lowP+lensing+non-CMB and 
TT+lowP+lensing+non-CMB+SPTpol-TE+EE, we again find that the non-CMB data are
significantly better at helping constrain cosmological model parameters 
than are the SPTpol-TE+EE data.

A summary of cosmological parameter values estimated from the three
different combination of data sets is given in Table
\ref{tab:para_nonflat_planck}. From this table and the figures, we see
that, in the case of the untilted non-flat dark energy models, the main effects
of adding the SPTpol TE+EE data to the TT+lowP+lensing+non-CMB data are a very
slight tightening of the constraints on $\theta_\textrm{MC}$ and $\Omega_b h^2$.

It is particularly interesting that including the SPTpol TE+EE data in a joint
analysis with either just the Planck 2015 TT+lowP+lensing data or with the
TT+lowP+lensing+non-CMB data still results in a detection of non-zero spatial
curvature with the TT+lowP+lensing+SPTpol-TE+EE data favoring a closed untilted model 
over the corresponding untilted flat limit at between $1.0\sigma$ and $\mathbf{1.6\sigma}$ and 
the TT+lowP+lensing+non-CMB+SPTpol-TE+EE data favoring a closed untilted model over the
corresponding untilted flat limit at between $3.1\sigma$ and $5.0\sigma$.

\begin{table*}
\caption{Untilted non-flat $\Lambda\textrm{CDM}$, XCDM, $\phi$CDM model parameters constrained by using the SPTpol TE+EE, TE, and EE data
         (mean and 68.3\% confidence limits).}
\begin{ruledtabular}
\begin{tabular}{lcccc}
\\[-1mm]                         & \multicolumn{4}{c}{Untilted non-flat $\Lambda$CDM ($\tau=0.112 \pm 0.012$)}              \\[+1mm]
\cline{2-5}\\[-1mm]
  Parameter                      & SPTpol TE+EE           & SPTpol TE              &  SPTpol EE            & TT+lowP+lensing+Non-CMB \\[+1mm]
 \hline \\[-1mm]
  $\Omega_b h^2$                 & $0.02280 \pm 0.00048$  & $0.02354 \pm 0.00084$  & $0.02281 \pm 0.00099$ & $0.02305 \pm 0.00019$   \\[+1mm]
  $\Omega_c h^2$                 & $0.1120 \pm 0.0054$    & $0.1185 \pm 0.0078$    & $0.1104 \pm 0.0090$   & $0.1093 \pm 0.0010$     \\[+1mm]
  $100\theta_\textrm{MC}$        & $1.0397 \pm 0.0013$    & $1.0395 \pm 0.0016$    & $1.0403 \pm 0.0015$   & $1.04227 \pm 0.00041$   \\[+1mm]
  $10^9 A_s e^{-2\tau}$          & $1.778 \pm 0.042$      & $1.835 \pm 0.066$      & $1.761 \pm 0.050$     & $1.831 \pm 0.010$       \\[+1mm]
  $\Omega_k$                     & $0.017 \pm 0.012$      & $-0.001 \pm 0.031$     & $0.014 \pm 0.018$     & $-0.0083 \pm 0.0016$     \\[+1mm]
 \hline \\[-1mm]
  $H_0$ [km s$^{-1}$ Mpc$^{-1}$] & $84.4 \pm 8.8$         & $73 \pm 13$            & $84 \pm 11$           & $68.01 \pm 0.62$        \\[+1mm]
  $\Omega_m$                     & $0.196 \pm 0.043$      & $0.30 \pm 0.12$        & $0.200 \pm 0.057$     & $0.2875 \pm 0.0055$     \\[+1mm]
  $\sigma_8$                     & $0.817 \pm 0.031$      & $0.845 \pm 0.045$      & $0.804 \pm 0.054$     & $0.8121 \pm 0.0095$     \\[+1mm]
 \hline \\[-1mm]
                                 & \multicolumn{4}{c}{Untilted non-flat XCDM ($\tau=0.119 \pm 0.012$)}              \\[+1mm]
\cline{2-5}\\[-1mm]
  Parameter                      & SPTpol TE+EE           & SPTpol TE              &  SPTpol EE            & TT+lowP+lensing+Non-CMB \\[+1mm]
 \hline \\[-1mm]
  $\Omega_b h^2$                 & $0.02271 \pm 0.00051$  & $0.02323 \pm 0.00098$  & $0.0222 \pm 0.0010$   & $0.02305 \pm 0.00020$   \\[+1mm]
  $\Omega_c h^2$                 & $0.1135 \pm 0.0060$    & $0.1199 \pm 0.0082$    & $0.117 \pm 0.011$     & $0.1092 \pm 0.0010$     \\[+1mm]
  $100\theta_\textrm{MC}$        & $1.0395 \pm 0.0013$    & $1.0395 \pm 0.0016$    & $1.0401 \pm 0.0015$   & $1.04227 \pm 0.00042$   \\[+1mm]
  $10^9 A_s e^{-2\tau}$          & $1.773 \pm 0.043$      & $1.815 \pm 0.073$      & $1.782 \pm 0.054$     & $1.832 \pm 0.010$       \\[+1mm]
  $\Omega_k$                     & $-0.016 \pm 0.063$     & $-0.008 \pm 0.061$     & $0.019 \pm 0.061$     & $-0.0069 \pm 0.0020$     \\[+1mm]
  $w$                            & $-0.46 \pm 0.33$       & $-0.61 \pm 0.50$       & $-0.42 \pm 0.28$      & $-0.960 \pm 0.032$     \\[+1mm]
 \hline \\[-1mm]
  $H_0$ [km s$^{-1}$ Mpc$^{-1}$] & $58 \pm 17$            & $59 \pm 16$            & $60 \pm 16$           & $67.45 \pm 0.75$        \\[+1mm]
  $\Omega_m$                     & $0.52 \pm 0.27$        & $0.50 \pm 0.25$        & $0.47 \pm 0.23$       & $0.2923 \pm 0.0066$     \\[+1mm]
  $\sigma_8$                     & $0.63 \pm 0.12$        & $0.70 \pm 0.16$        & $0.60 \pm 0.11$       & $0.805 \pm 0.011$     \\[+1mm]
 \hline \\[-1mm]
                                 & \multicolumn{4}{c}{Untilted non-flat $\phi$CDM ($\tau=0.122 \pm 0.012$)}              \\[+1mm]
\cline{2-5}\\[-1mm]
  Parameter                      & SPTpol TE+EE           & SPTpol TE              &  SPTpol EE            & TT+lowP+lensing+Non-CMB \\[+1mm]
 \hline \\[-1mm]
  $\Omega_b h^2$                 & $0.02263 \pm 0.00048$  & $0.02280 \pm 0.00095$  & $0.02210 \pm 0.00096$ & $0.02304 \pm 0.00020$   \\[+1mm]
  $\Omega_c h^2$                 & $0.1139 \pm 0.0057$    & $0.1220 \pm 0.0087$    & $0.118 \pm 0.010$     & $0.1093 \pm 0.0010$     \\[+1mm]
  $H_0$ [km s$^{-1}$ Mpc$^{-1}$] & $57 \pm 13$            & $61 \pm 15$            & $63 \pm 14$           & $67.36 \pm 0.72$        \\[+1mm]
  $10^9 A_s e^{-2\tau}$          & $1.776 \pm 0.039$      & $1.790 \pm 0.070$      & $1.790 \pm 0.055$     & $1.832 \pm 0.010$       \\[+1mm]
  $\Omega_k$                     & $0.010 \pm 0.036$      & $0.031 \pm 0.058$      & $0.043 \pm 0.049$     & $-0.0063 \pm 0.0020$     \\[+1mm]
  $\alpha$                       & $4.3 \pm 2.7$          & $ < 9.7$ [95.4\% C.L.] & $4.8 \pm 2.7$         & $ < 0.31$ [95.4\% C.L.]  \\[+1mm]
 \hline \\[-1mm]
  $100\theta_\textrm{MC}$        & $1.0394 \pm 0.0013$    & $1.0391 \pm 0.0016$    & $1.0399 \pm 0.0015$   & $1.04210 \pm 0.00041$   \\[+1mm]
  $\Omega_m$                     & $0.47 \pm 0.17$        & $0.46 \pm 0.19$        & $0.41 \pm 0.16$       & $0.2931 \pm 0.0064$     \\[+1mm]
  $\sigma_8$                     & $0.619 \pm 0.075$      & $0.63 \pm 0.11$        & $0.600 \pm 0.081$     & $0.805 \pm 0.011$     \\[+1mm]
\end{tabular}
\\[+1mm]
\begin{flushleft}
Note: Parameter constraints for Planck 2015 TT+lowP+lensing and non-CMB (SN, BAO, $H(z)$, $f\sigma_8$) data sets are from Ref.\ \cite{ParkRatra2018b} for the $\Lambda$CDM and XCDM models and from Ref.\ \cite{ParkRatra2018c} for the $\phi$CDM model. For the SPTpol analyses, a different Gaussian prior for $\tau$ (indicated in the subheadings) has been used for each cosmological model (see main text for discussions and details).
\end{flushleft}
\end{ruledtabular}
\label{tab:para_nonflat}
\end{table*}

We next examine the consistency between the untilted non-flat model constraints
obtained by using the Planck CMB and non-CMB data and those determined from the 
SPTpol data sets. The results are summarized in Table \ref{tab:chi2_nonflat}. 
Compared to the tilted flat cases, the best-fit untilted non-flat models 
favored by the Planck CMB and non-CMB data have similar 
$\chi_\textrm{min}^2$ and $N_\sigma$ values.  For the best-fit models favored 
by the SPTpol TE+EE and EE data, the non-flat models improve data fitting 
with smaller $\chi_\textrm{min}^2$ and $N_\sigma$ values than the flat models 
except for the case of the XCDM parameterization with SPTpol TE+EE data 
\cite{DIC}.
On the other hand, the best-fit untilted non-flat dark energy models
all poorly fit the SPTpol TE data with larger $\chi_\textrm{min}^2$ values, 
compared to
the corresponding tilted flat models. The increase in the $\chi_\textrm{min}^2$
is very notable in the case of the best-fit non-flat XCDM model favored by the
SPTpol TE data.
For SPTpol TE+EE data, the $N_\sigma$ values are all larger than $2.2$
so none of the best-fit models provide a good fit to the SPTpol TE+EE data.
Comparing the $\chi_p^2$ values of Table \ref{tab:chi2_nonflat}
for the untilted non-flat $\Lambda$CDM, XCDM, and $\phi$CDM models, we see
that there is also no significant evidence of tension between the non-flat dark energy
models constrained using the Planck CMB and non-CMB data and that
constrained using SPTpol data alone.

Except for the untilted non-flat $\Lambda$CDM model where the SPTpol TE+EE
data favors a $\sigma_8$ value ($\sigma_8=0.817\pm 0.031$) that is 0.5$\sigma$ 
larger than that favored by the Planck 2015 TT+lowP+lensing data 
($\sigma_8=0.799 \pm 0.021$, \cite{ParkRatra2018a}),
the SPTpol TE+EE data favor $\sigma_8$ values 
between 1.1$\sigma$ and 2.7$\sigma$ lower
than what Planck does.
This has been noted for the tilted flat $\Lambda$CDM model in Ref.\
\cite{Henningetal2018}. By comparing the blue and red contours in the
$\sigma_8$--$\Omega_m$ panels of Figs.\
\ref{fig:like_FL_P2015}--\ref{fig:like_FQ_P2015} and Figs.\
\ref{fig:like_NL_SPT_planck}--\ref{fig:like_NQ_SPT_planck} (even in Fig.\
\ref{fig:like_NL_SPT_planck} for the untilted non-flat $\Lambda$CDM model) we see
that adding the SPTpol TE+EE data to the mix results in a slight shift of the
contours in the direction that eases tension with weak lensing measurements.
We emphasize however that this shift is not as significant as that caused
by non-zero spatial curvature in the closed models, see Refs.\ \cite{Oobaetal2018a, Oobaetal2018b, Oobaetal2018c, ParkRatra2018a,  ParkRatra2018b,  ParkRatra2018c}.

While the SPTpol data $H_0$ value is larger than the Planck 2015 CMB and 
non-CMB data $H_0$ value in the tilted flat $\Lambda$CDM model (first seen 
in Ref.\ \cite{Henningetal2018}), this is not true in most of the other models.
Additionally, adding the SPTpol data to the Planck 2015 CMB and
non-CMB data does not result in a significant change in the measured $H_0$'s, 
which are quite consistent with a number of recent $H_0$ estimates, \cite{HubbleConstant}.

\begin{figure*}[htbp]
\centering
\mbox{\includegraphics[width=180mm, trim={30 475 50 70},clip]{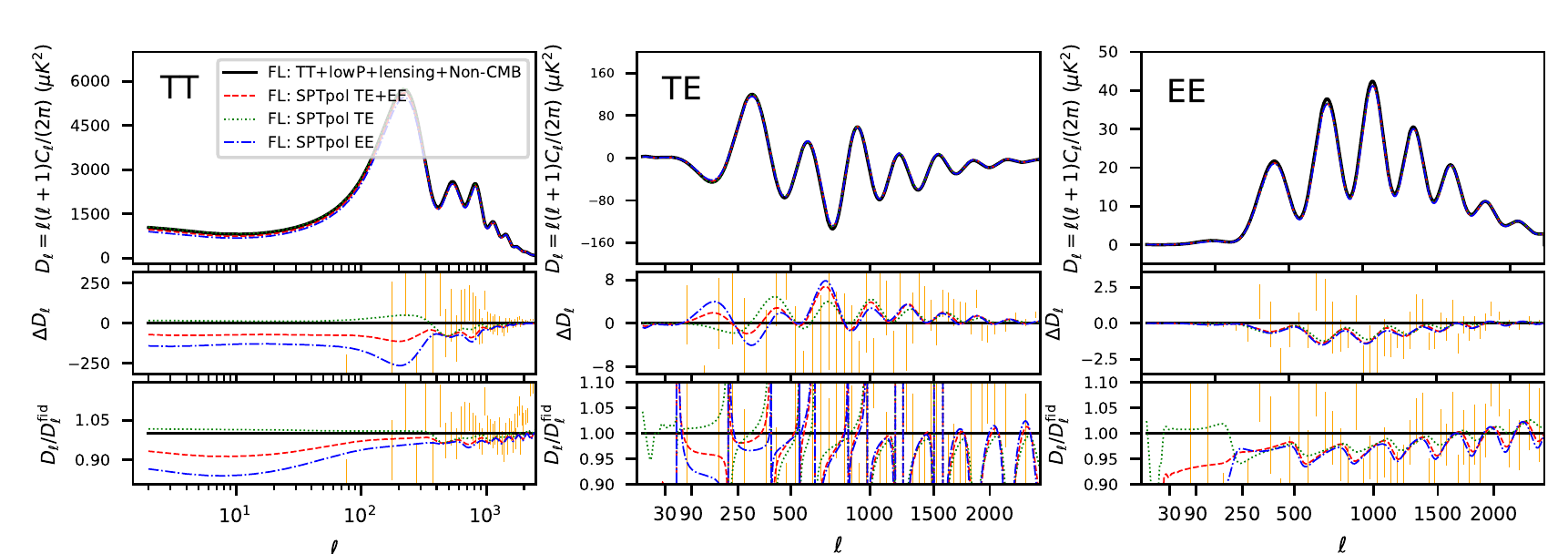}}\\
\mbox{\includegraphics[width=180mm, trim={30 475 50 70},clip]{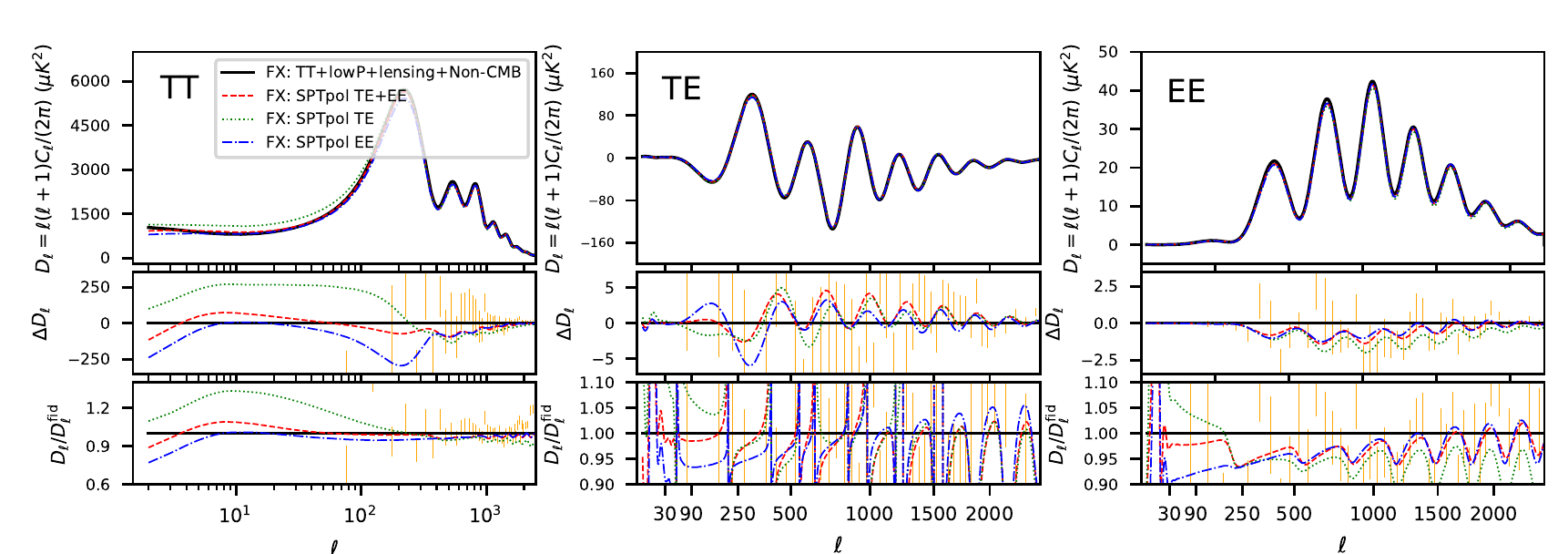}}\\
\mbox{\includegraphics[width=180mm, trim={30 475 50 70},clip]{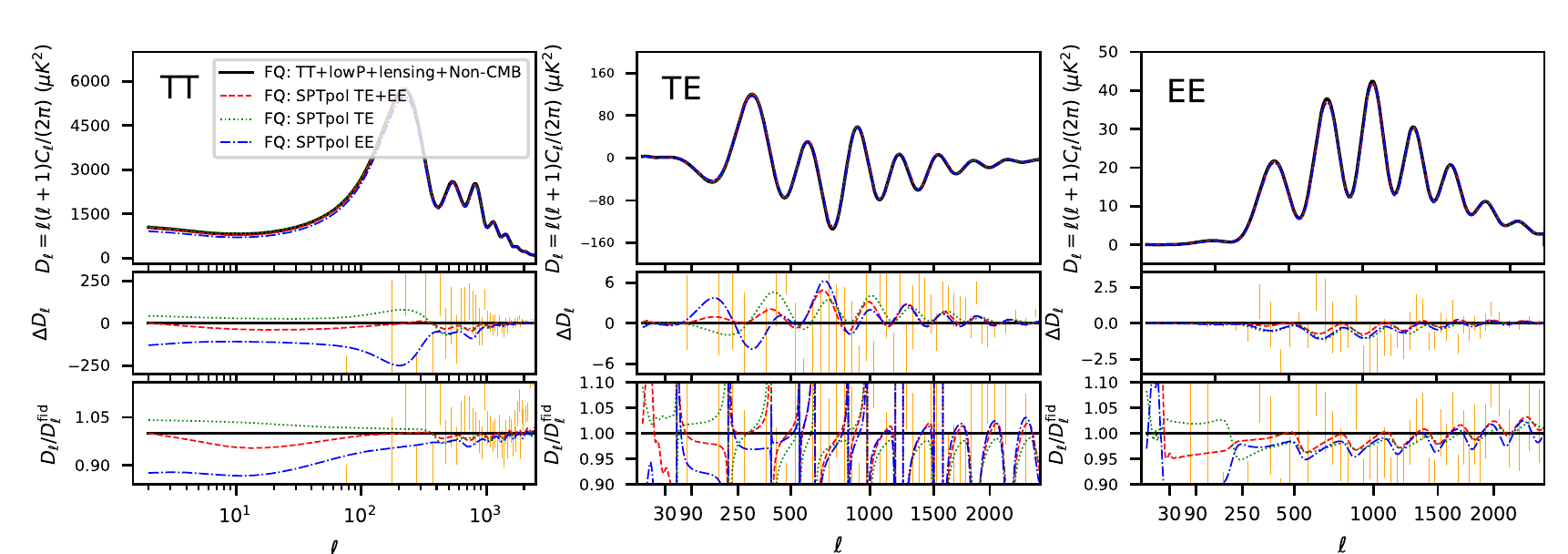}}
\caption{CMB power spectra of best-fit tilted flat $\Lambda$CDM (FL, upper 
         row), XCDM (FX, middle row), and $\phi$CDM (FQ, lower row)
         models constrained by using TT+lowP+lensing+non-CMB data, and
         SPTpol TE+EE, TE, and EE data sets. Difference ($\Delta D_{\ell}$) 
         and ratio ($D_{\ell} / D_{\ell}^{\textrm{fid}}$) panels show quantities 
         with respect to the fiducial model constrained using 
         TT+lowP+lensing+non-CMB data. Vertical error bars indicate the 
         confidence limits of the SPTpol power spectrum data including TT band powers. 
}
\label{fig:cl_FLXQ}
\end{figure*}

\begin{figure*}[htbp]
\centering
\mbox{\includegraphics[width=180mm, trim={30 475 50 70},clip]{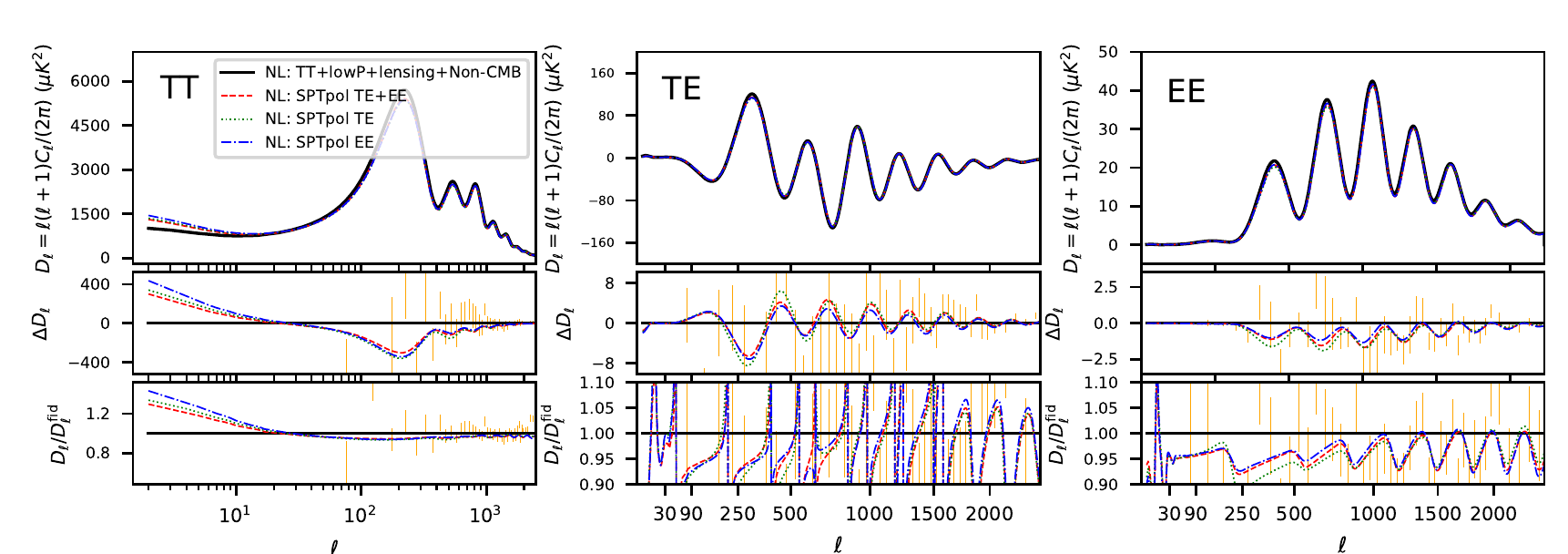}}\\
\mbox{\includegraphics[width=180mm, trim={30 475 50 70},clip]{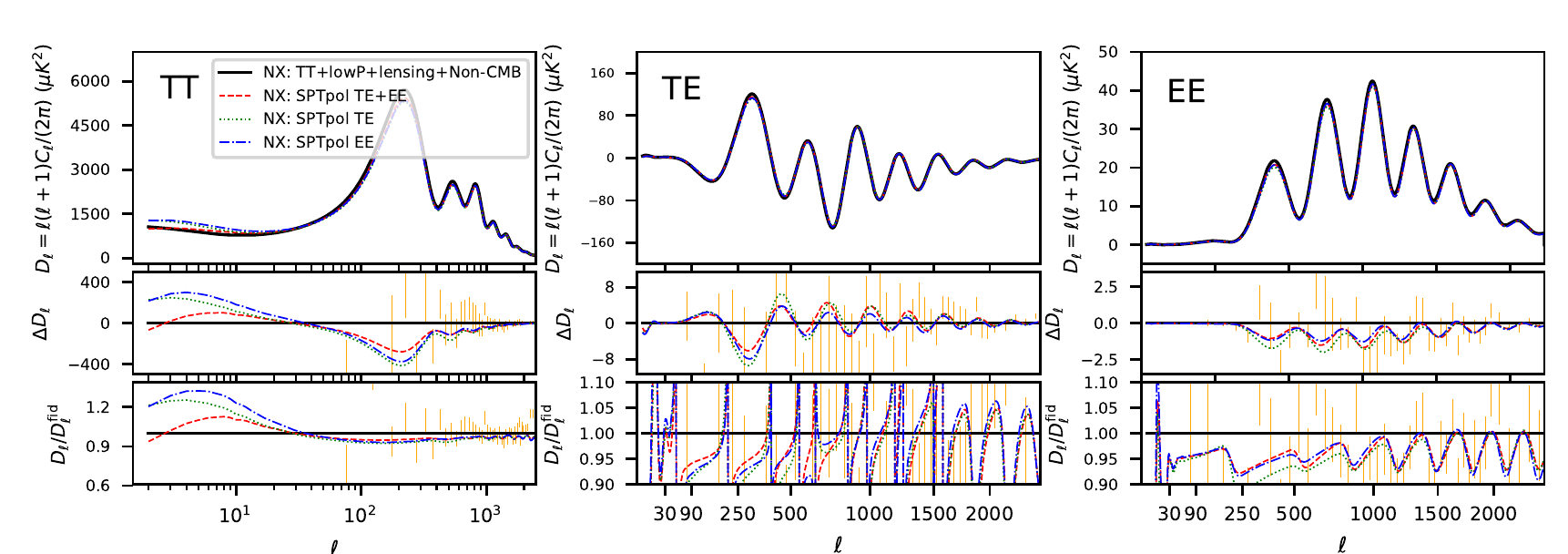}}\\
\mbox{\includegraphics[width=180mm, trim={30 475 50 70},clip]{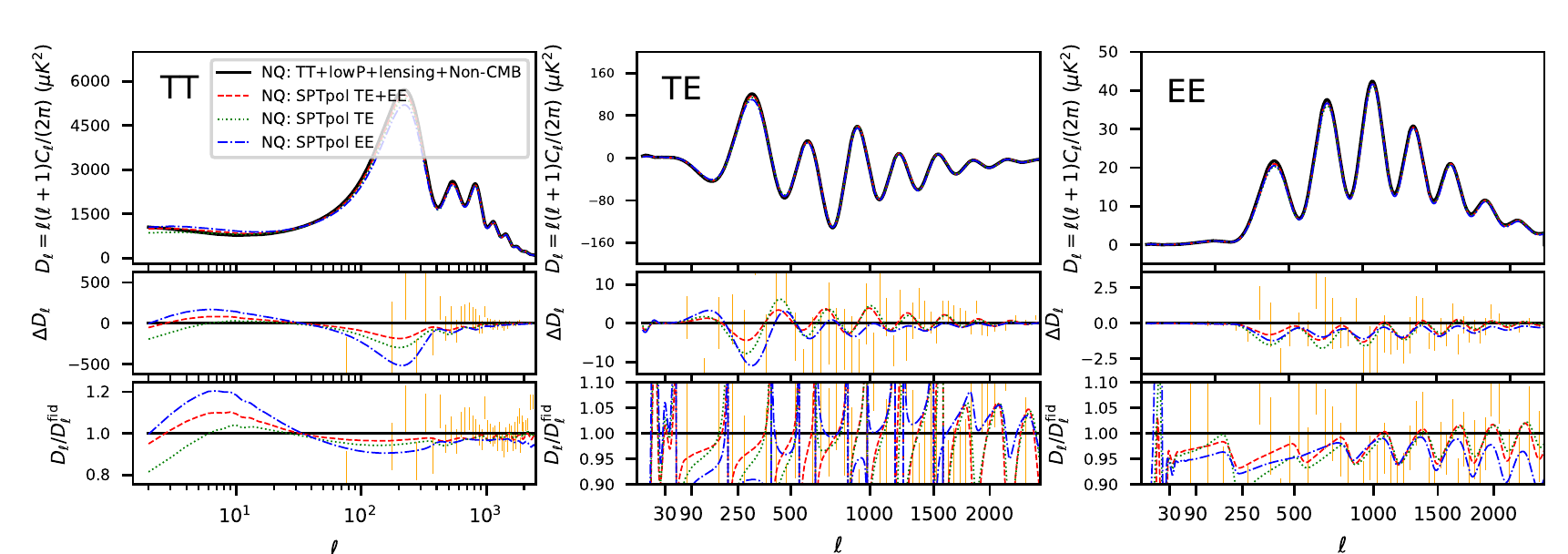}}
\caption{CMB power spectra of best-fit untilted non-flat $\Lambda$CDM (NL, upper
         row), XCDM (NX, middle row), and $\phi$CDM (NQ, lower row)
         models constrained by using TT+lowP+lensing+non-CMB data, and SPTpol 
         TE+EE, TE, and EE data sets. Difference ($\Delta D_{\ell}$) and ratio 
         ($D_{\ell} / D_{\ell}^{\textrm{fid}}$) panels show quantities with 
         respect to the fiducial model constrained using 
         TT+lowP+lensing+non-CMB data. Vertical error bars indicate the 
         confidence limits of the SPTpol power spectrum data including TT band powers. 
}
\label{fig:cl_NLXQ}
\end{figure*}

While the PTEs in Tables \ref{tab:chi2_flat} and \ref{tab:chi2_nonflat}
indicate no significant evidence for tension between models constrained using
Planck 2015 CMB and non-CMB data and constrained using the SPTpol TE+EE data, 
the $N_\sigma$
values are always larger than 2.2 and sometimes exceed 3. This means that
CMB power spectra of the models that best fit the Planck 2018 data, 
the TT+lowP+lensing+non-CMB data, and the SPTpol TE+EE data do not provide
good fits to the SPTpol TE+EE data. This is clearly illustrated in 
Figs. \ref{fig:cl_FLXQ}--\ref{fig:cl_NLXQ} that show the TT, TE, and EE power 
spectra that best fit the SPTpol TE+EE, TE, and EE data sets in the various 
dark energy models, together with the difference and ratio of power spectrum 
with respect to the best-fit fiducial model constrained by using the Planck 
2015 CMB and non-CMB data. From these figures we see that the best-fit 
models favored by the SPTpol data do not fit well the CMB temperature power 
spectrum in the low $\ell$ range ($\ell < 200$). 

\begin{table*}
\caption{Untilted non-flat $\Lambda\textrm{CDM}$, XCDM, $\phi$CDM model parameters constrained by using Planck 2015, SPTpol TE+EE, and non-CMB data
         (mean and 68.3\% confidence limits).}
\begin{ruledtabular}
\begin{tabular}{lccc}
\\[-1mm]                         & \multicolumn{3}{c}{Untilted non-flat $\Lambda$CDM}              \\[+1mm]
\cline{2-4}\\[-1mm]
  Parameter                      & TT+lowP+lensing+SPTpol & TT+lowP+lensing+non-CMB & TT+lowP+lensing+non-CMB+SPTpol  \\[+1mm]
 \hline \\[-1mm]
  $\Omega_b h^2$                 & $0.02303 \pm 0.00018$  & $0.02305 \pm 0.00019$   & $0.02303 \pm 0.00018$\\[+1mm]
  $\Omega_c h^2$                 & $0.1091 \pm 0.0011$    & $0.1093 \pm 0.0010$     & $0.1093 \pm 0.0010$\\[+1mm]
  $100\theta_\textrm{MC}$        & $1.04206 \pm 0.00039$  & $1.04227 \pm 0.00041$   & $1.04201 \pm 0.00039$\\[+1mm]
  $\tau$                         & $0.100 \pm 0.021$      & $0.112 \pm 0.012$       & $0.108 \pm 0.011$\\[+1mm]
  $\Omega_k$                     & $-0.0139 \pm 0.0086$   & $-0.0083 \pm 0.0016$    & $-0.0080 \pm 0.0016$ \\[+1mm]
  $\ln(10^{10} A_s)$             & $3.107 \pm 0.042$      & $3.132 \pm 0.022$       & $3.123 \pm 0.021$ \\[+1mm]
 \hline \\[-1mm]
  $10^9 A_s e^{-2\tau}$          & $1.829 \pm 0.010$      & $1.831 \pm 0.010$       & $1.8292 \pm 0.0100$\\[+1mm]
  $H_0$ [km s$^{-1}$ Mpc$^{-1}$] & $65.8 \pm 3.4$         & $68.01 \pm 0.62$        & $68.04 \pm 0.61$\\[+1mm]
  $\Omega_m$                     & $0.309 \pm 0.032$      & $0.2875 \pm 0.0055$     & $0.2872 \pm 0.0052$\\[+1mm]
  $\sigma_8$                     & $0.799 \pm 0.021$      & $0.8121 \pm 0.0095$     & $0.8081 \pm 0.0093$\\[+1mm]
 \hline \\[-1mm]
                                 & \multicolumn{3}{c}{Untilted non-flat XCDM}              \\[+1mm]
\cline{2-4}\\[-1mm]
  Parameter                      & TT+lowP+lensing+SPTpol & TT+lowP+lensing+non-CMB & TT+lowP+lensing+non-CMB+SPTpol \\[+1mm]
 \hline \\[-1mm]
  $\Omega_b h^2$                 & $0.02304 \pm 0.00018$  & $0.02305 \pm 0.00020$   & $0.02304 \pm 0.00018$ \\[+1mm]
  $\Omega_c h^2$                 & $0.1091 \pm 0.0011$    & $0.1092 \pm 0.0010$     & $0.1092 \pm 0.0010$ \\[+1mm]
  $100\theta_\textrm{MC}$        & $1.04208 \pm 0.00039$  & $1.04227 \pm 0.00042$   & $1.04199 \pm 0.00039$ \\[+1mm]
  $\tau$                         & $0.099 \pm 0.021$      & $0.119 \pm 0.012$       & $0.115 \pm 0.012$\\[+1mm]
  $\Omega_k$                     & $-0.018 \pm 0.018$     & $-0.0069 \pm 0.0020$    & $-0.0066 \pm 0.0019$  \\[+1mm]
  $w$                            & $-1.08 \pm 0.40$       & $-0.960 \pm 0.032$      & $-0.958 \pm 0.032$ \\[+1mm]
  $\ln(10^{10} A_s)$             & $3.104 \pm 0.042$      & $3.146 \pm 0.024$       & $3.137 \pm 0.024$ \\[+1mm]
 \hline \\[-1mm]
  $10^9 A_s e^{-2\tau}$          & $1.829 \pm 0.010$      & $1.832 \pm 0.010$       & $1.830 \pm 0.010$ \\[+1mm]
  $H_0$ [km s$^{-1}$ Mpc$^{-1}$] & $69 \pm 15$            & $67.45 \pm 0.75$        & $67.46 \pm 0.76$ \\[+1mm]
  $\Omega_m$                     & $0.32 \pm 0.14$        & $0.2923 \pm 0.0066$     & $0.2921 \pm 0.0067$ \\[+1mm]
  $\sigma_8$                     & $0.82 \pm 0.11$        & $0.805 \pm 0.011$       & $0.801 \pm 0.011$ \\[+1mm]
 \hline \\[-1mm]
                                 & \multicolumn{3}{c}{Untilted non-flat $\phi$CDM}              \\[+1mm]
\cline{2-4}\\[-1mm]
  Parameter                      & TT+lowP+lensing+SPTpol & TT+lowP+lensing+non-CMB & TT+lowP+lensing+non-CMB+SPTpol \\[+1mm]
 \hline \\[-1mm]
  $\Omega_b h^2$                 & $0.02302 \pm 0.00018$  & $0.02304 \pm 0.00020$   & $0.02304 \pm 0.00018$ \\[+1mm]
  $\Omega_c h^2$                 & $0.1091 \pm 0.0010$    & $0.1093 \pm 0.0010$     & $0.1092 \pm 0.0010$  \\[+1mm]
  $H_0$ [km s$^{-1}$ Mpc$^{-1}$] & $54.7 \pm 6.9$         & $67.36 \pm 0.72$        & $67.36 \pm 0.72$   \\[+1mm]
  $\tau$                         & $0.102 \pm 0.020$      & $0.122 \pm 0.012$       & $0.117 \pm 0.012$  \\[+1mm]
  $\Omega_k$                     & $-0.034 \pm 0.021$     & $-0.0063 \pm 0.0020$    & $-0.0061 \pm 0.0020$  \\[+1mm]
  $\alpha$ [95.4\% C.L.]         & $ < 4.6$              & $ < 0.31$               & $ < 0.32$  \\[+1mm]
 \hline \\[-1mm]
  $10^9 A_s e^{-2\tau}$          & $1.828 \pm 0.010$      & $1.832 \pm 0.010$       & $1.830 \pm 0.010$ \\[+1mm]
  $100\theta_\textrm{MC}$        & $1.04186 \pm 0.00039$  & $1.04210 \pm 0.00041$   & $1.04183 \pm 0.00039$ \\[+1mm]
  $\Omega_m$                     & $0.47 \pm 0.12$        & $0.2931 \pm 0.0064$     & $0.2930 \pm 0.0064$ \\[+1mm]
  $\sigma_8$                     & $0.707 \pm 0.056$      & $0.805 \pm 0.011$       & $0.800 \pm 0.010$ \\[+1mm]
\end{tabular}
\\[+1mm]
\begin{flushleft}
Note: Parameter constraints for Planck 2015 TT+lowP+lensing and non-CMB (SN, BAO, $H(z)$, $f\sigma_8$) data sets are from Refs.\ \cite{ParkRatra2018b} for the $\Lambda$CDM and XCDM models and from Ref.\ \cite{ParkRatra2018c} for the $\phi$CDM model.
\end{flushleft}
\end{ruledtabular}
\label{tab:para_nonflat_planck}
\end{table*}

\begin{table*}
\caption{Minimum $\chi^2$ values for the SPTpol TE+EE, TE, EE spectra in the best-fit untilted
         non-flat $\Lambda$CDM, XCDM, and $\phi$CDM models constrained by using Planck 2015 TT+lowP+SN+BAO+$H(z)$+$f\sigma_8$ data \cite{ParkRatra2018b,ParkRatra2018c}. }
\begin{ruledtabular}
\begin{tabular}{lrcccccc}
                       &       & \multicolumn{2}{c}{Best-fit untilted non-flat $\Lambda$CDM}     &  \multicolumn{2}{c}{SPT ($\Lambda$CDM $\tau$ \cite{ParkRatra2018b})} &   &  \\[+1mm]
\cline{3-4}\cline{5-6}\\[-1mm]
    SPTpol spectrum    & $N_b$ & $\chi_{\textrm{min}}^2$  & $N_{\sigma}$  & $\chi_{\textrm{min}}^2$  & $N_{\sigma}$ & $\chi_p^2$   &  PTE   \\[+1mm]
\hline\\[-1mm]
    TE + EE            & 112   & 146.96                   & 2.98          & 136.18                   & 2.23          &   10.31     & 0.067       \\[+1mm]
    TE                 &  56   &  73.88                   & 2.64          &  68.72                   & 2.11          &    4.09     & 0.536       \\[+1mm]
    EE                 &  56   &  68.46                   & 2.09          &  59.05                   & 1.13          &    6.73     & 0.242       \\[+1mm]
\hline\\[-1mm]
                       &       & \multicolumn{2}{c}{Best-fit untilted non-flat XCDM}     &  \multicolumn{2}{c}{SPT (XCDM $\tau$ \cite{ParkRatra2018b})}   &             &  \\[+1mm]
\cline{3-4}\cline{5-6}\\[-1mm]
    SPTpol spectrum    & $N_b$ & $\chi_{\textrm{min}}^2$  & $N_{\sigma}$  & $\chi_{\textrm{min}}^2$  & $N_{\sigma}$  & $\chi_p^2$  &  PTE   \\[+1mm]
\hline\\[-1mm]
    TE + EE            & 112   & 147.46                   & 3.10          & 136.45                   & 2.33          &    8.24     & 0.221       \\[+1mm]
    TE                 &  56   &  74.14                   & 2.80          &  68.81                   & 2.25          &    4.34     & 0.630      \\[+1mm]
    EE                 &  56   &  68.78                   & 2.25          &  58.85                   & 1.22          &    9.01     & 0.173       \\[+0mm]
\hline\\[-1mm]
                       &       & \multicolumn{2}{c}{Best-fit untilted non-flat $\phi$CDM}   &  \multicolumn{2}{c}{SPT ($\phi$CDM $\tau$ \cite{ParkRatra2018c})}  &         &  \\[+1mm]
\cline{3-4}\cline{5-6}\\[-1mm]
    SPTpol spectrum    & $N_b$ & $\chi_{\textrm{min}}^2$  & $N_{\sigma}$  & $\chi_{\textrm{min}}^2$  & $N_{\sigma}$  & $\chi_p^2$  &  PTE ($H_0$)  \\[+1mm]
\hline\\[-1mm]
    TE + EE            & 112   & 146.81                   & 3.05          & 137.34                   & 2.39          &    6.45     & 0.375  \\[+1mm]
    TE                 &  56   &  73.80                   & 2.76          &  69.32                   & 2.30          &    5.30     & 0.506  \\[+1mm]
    EE                 &  56   &  68.43                   & 2.21          &  60.22                   & 1.36          &    7.58     & 0.270   \\[+1mm]
\cline{3-4}\cline{5-6}\\[-1mm]
    SPTpol spectrum    & $N_b$ & $\chi_{\textrm{min}}^2$  & $N_{\sigma}$  & $\chi_{\textrm{min}}^2$  & $N_{\sigma}$  & $\chi_p^2$  &  PTE ($\theta_\textrm{MC}$)  \\[+1mm]
\hline\\[-1mm]
    TE + EE            & 112   & 146.81                   & 3.05          & 137.34                   & 2.39          &   10.19     & 0.117  \\[+1mm]
    TE                 &  56   &  73.80                   & 2.76          &  69.32                   & 2.30          &    6.93     & 0.328  \\[+1mm]
    EE                 &  56   &  68.43                   & 2.21          &  60.22                   & 1.36          &    8.88     & 0.181   \\[+0mm]
\end{tabular}
\\[+1mm]
\begin{flushleft}
Note: We assume a different Gaussian prior of $\tau$ for each cosmological model. For our best-fit untilted non-flat $\Lambda$CDM, XCDM, and $\phi$CDM models,
we apply $\tau=0.112 \pm 0.012$, $0.119 \pm 0.012$, and $0.122 \pm 0.012$ \cite{ParkRatra2018b,ParkRatra2018c}, respectively.
\end{flushleft}
\end{ruledtabular}
\label{tab:chi2_nonflat}
\end{table*}

\section{Summary}

We have constrained tilted flat and untilted non-flat dark energy inflation
models by using SPTpol CMB, Planck 2015 CMB, and non-CMB data. In each case we have considered three different dark energy models, a time-independent cosmological constant as well as ideal $X$-fluid and scalar field dynamical dark energy density models.

In summary, our main findings are:
\begin{itemize}
\item All the Planck CMB and non-CMB data best-fit model CMB anisotropy power spectra we consider do not provide great fits to the SPTpol TE+EE data.
Models that best-fit the Planck CMB and non-CMB data give larger minimum $\chi^2$ when they try to fit the SPTpol data.  
\item In all models we consider, there is, however, no significant evidence 
of tension between a model constrained using Planck 2015 CMB and non-CMB data
and the same model constrained using SPTpol TE+EE data (i.e., given the uncertainities, in each model the cosmological parameter constraints from both sets of data are largely consistent with each other) and so it is appropriate to use these data together to jointly constrain model parameters.
(We note that Ref.\ \cite{Henningetal2018} found some tension between the 
tilted flat $\Lambda$CDM model constrained using just the Planck 2015 data and 
that constrained using the SPTpol TE+EE data.)
\item Depending on cosmological model, the SPTpol TE+EE data can favor a larger
or smaller $H_0$ than is favored by the Planck 2015 data.
\item In most models the SPTpol TE+EE data favor a lower $\sigma_8$ than is 
favored by the Planck 2015 data and moves the $\sigma_8$--$\Omega_m$ Planck 
2015 CMB and 
non-CMB data contours in the direction of reducing tension with weak lensing 
measurements, but the overall effect is very small.   
\item When the smaller angular scale SPTpol TE+EE data is used to jointly 
analyze untilted non-flat models with the TT+lowP+lensing data or with the 
TT+lowP+lensing+non-CMB data, closed untilted models with non-zero $\Omega_k$ are 
still favored over the corresponding $\Omega_k = 0$ untilted cases.
\end{itemize} 

Combined with the Planck CMB data, the non-CMB data, especially the BAO data, constrain the cosmological model parameters significantly better than do the SPTpol TE+EE data.
While near-future 
ground-based CMB anisotropy experiments will produce data with better 
constraining power, perhaps data from a future space-based experiment might 
be more helpful for this purpose.

Recently, there was a report that the Planck 2018 CMB spectra prefer a positive curvature at more than 99\% confidence level \cite{DiValentinoetal2019}. According to the tilted nonflat $\Lambda$CDM model constraints of Ref.\ \cite{DiValentinoetal2019}, a quite low value of Hubble constant is favored and $\Omega_m$ and $\sigma_8$ parameters constrained by the Planck CMB data have discordance with the cosmic shear measurements. On the other hand, our untilted nonflat dark energy models have a Hubble constant that is very similar to that of the flat $\Lambda$CDM model and give $\Omega_m$ and $\sigma_8$ values that are consistent with those from the weak lensing observations (see Refs.\ \cite{ParkRatra2018a,ParkRatra2018b,ParkRatra2018c}).

It has been known that there are systematic difference between model parameters obtained with high $\ell$ and low $\ell$ Planck 2015 data \cite{Addisonetal2016,Aghanimetal2017}. In addition, the South Pole Telescope team reported that there might be inconsistencies between the SPT CMB temperature anisotropy and the Planck 2015 high $\ell$ data in the tilted flat $\Lambda$CDM model \cite{Ayloretal2017}. Such tension between parameter constraints for the Planck and SPTpol data sets is reduced when the gravitational lensing scaling parameter $A_L$ is introduced as a free parameter \cite{Henningetal2018}. Whether these tensions persist in other dark energy models will be investigated in a subsequent study.  

%
%
\acknowledgements
C.-G.P.\ was supported by the Basic Science Research Program through the National Research Foundation of Korea (NRF)
funded by the Ministry of Education (No.\ 2017R1D1A1B03028384). B.R.\ was supported in part by DOE
grant DE-SC0019038.

\def\and{{and }}


\end{document}